%% file: main.tex
\documentclass[journal]{IEEEtran}

\ifCLASSOPTIONcompsoc
\else
\fi
\ifCLASSINFOpdf
\else
\fi
\usepackage{simplemargins}
\usepackage{graphicx}
\usepackage{indentfirst}
\usepackage{epsfig}
\usepackage{subfigure}
\usepackage{amsfonts}
\usepackage{color}
\usepackage{amsmath}
\usepackage{nomencl}

\usepackage{amssymb}
\usepackage{multicol}
\usepackage{rotating}
\usepackage{url}
\usepackage{bm}
\usepackage[ruled,commentsnumbered]{algorithm2e}
\usepackage{amsthm}
\usepackage{algpseudocode}
\usepackage[tikz]{bclogo}

\newcommand{\be}{\begin{itemize}} \newcommand{\ee}{\end{itemize}}
 
\newcommand{\tit}{\textit}

\makeatletter
\newcommand\figcaption{\def\@captype{figure}\caption}
\newcommand\tabcaption{\def\@captype{table}\caption}
\makeatother
\IEEEoverridecommandlockouts 

\begin{document}

\title{The Role of Caching in Future Communication Systems and Networks}

\author{\IEEEauthorblockN{ Georgios S. Paschos, \emph{Senior Member, IEEE}, George Iosifidis, Meixia Tao, \emph{Senior Member, IEEE}, Don Towsley, \emph{Fellow, IEEE}, Giuseppe Caire, \emph{Fellow, IEEE}}
\IEEEauthorblockA{}
\IEEEcompsocitemizethanks{
\IEEEcompsocthanksitem

G. Paschos is with France Research Center, Huawei Technologies, Paris (email: georgios.paschos@huawei.com).

G. Iosifidis is with the School of Computer Science and Statistics, Trinity College Dublin, Dublin 2, Ireland (e-mail: george.iosifidis@tcd.ie).

M. Tao is with the Department of Electronic Engineering, Shanghai Jiao Tong University, Shanghai 200240, China (email: mxtao@sjtu.edu.cn).

D. Towsley is with the School of Computer Science, University of Massachusetts Amherst, MA 01002, USA (email: towsley@cs.umass.edu).

G. Caire is with the Department of Electrical Engineering and Computer Science, Technical University of Berlin, 10623 Berlin, Germany (email: caire@tu-berlin.de).
}
}

\IEEEcompsoctitleabstractindextext{

\begin{abstract}
This paper has the following ambitious goal: to convince the reader that \emph{content caching} is an exciting research topic for the future communication systems and networks. Caching has been studied for more than 40 years, and has recently received increased attention from industry and academia. Novel caching techniques promise to push the network performance to unprecedented limits, but also pose significant technical challenges. This tutorial provides a brief overview of existing caching solutions, discusses seminal papers that open new directions in caching, and presents the contributions of this Special Issue. We analyze the challenges that caching needs to address today, considering also an industry perspective, and identify bottleneck issues that must be resolved to unleash the full potential of this promising technique.

\end{abstract}

}

\maketitle

\renewcommand{\thefootnote}{\noindent{}}
\renewcommand{\thefootnote}{\arabic{footnote}~}

\IEEEdisplaynotcompsoctitleabstractindextext
\IEEEpeerreviewmaketitle

\input{Section-IntroVer12}

\input{Section-Overview-1Ver12}

\input{Section-Overview-2Ver12}

\input{Section-Open-Issues-1Ver12}

\input{Section-Open-Issues-2Ver12}

\input{Section-ConclusionsVer12}

\section{Acknowledgments}
George Iosifidis acknowledges the support of Science Foundation Ireland, under grant 17/CDA/4760. The work of Don Towsley was sponsored by the U.S. ARL and the U.K. MoD under Agreement Number W911NF-16-3-0001 and by the NSF under grant NSF CNS-1617437. The work of Meixia Tao is supported by the National Natural Science Foundation of China under grants 61571299 and 61521062. The opinions expressed in this paper are of the authors alone, and do not represent an official position of Huawei Technologies. 

The authors would like to acknowledge the excellent work of all reviewers who participated in this double JSAC Issue; and the great support they received from Max Henrique Machado Costa, JSAC Senior Editor, and Laurel Greenidge, Executive Editor.

\input{ReferencesVer13}

%

\begin{IEEEbiography}[{\includegraphics[width=1.1in,height=1.45in,clip,keepaspectratio]{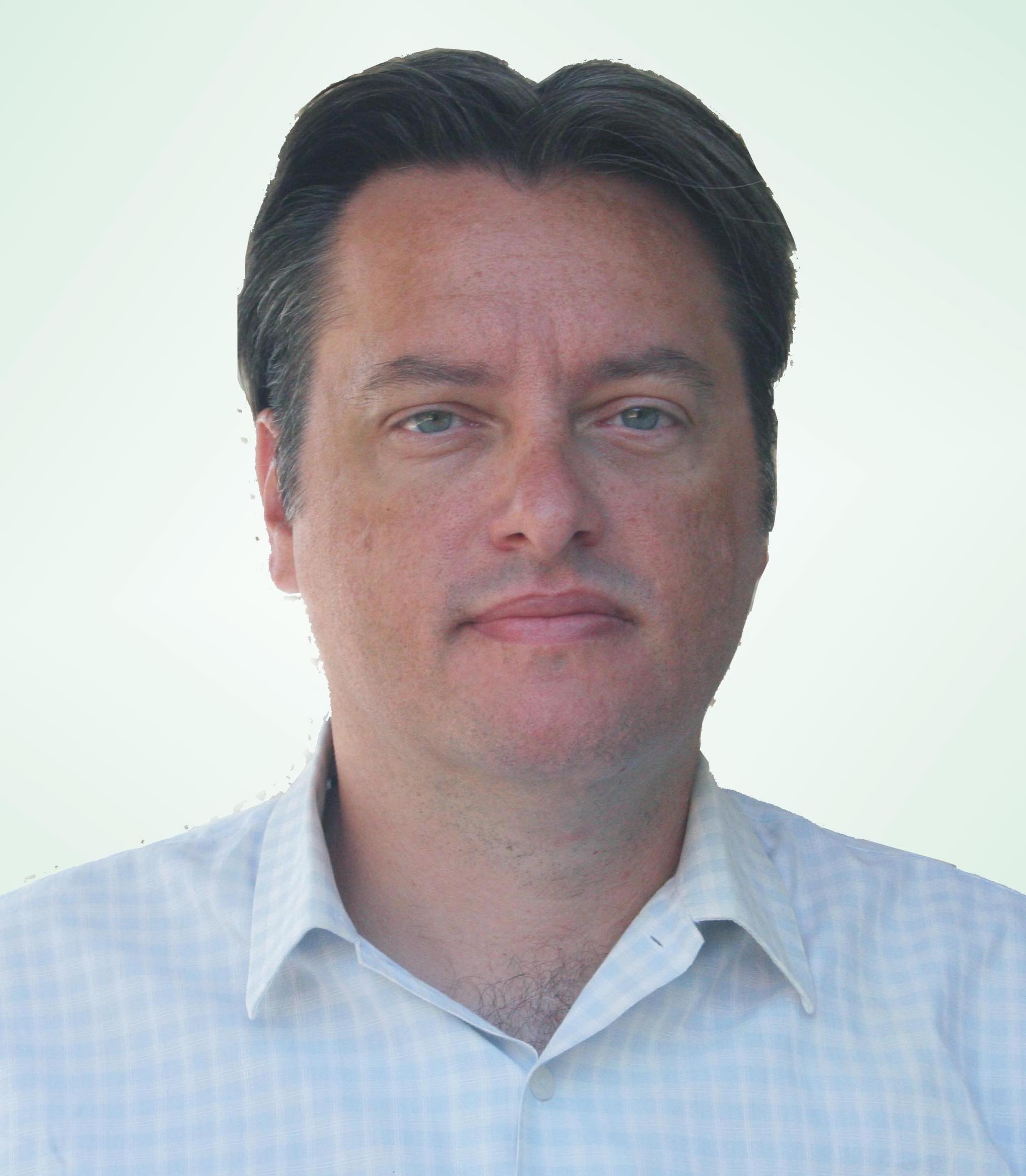}}]
{Georgios S. Paschos} is a principal researcher at Huawei Technologies, Paris, France, leading the Network Control and Resource Allocation team since Nov. 2014. Previously, he held research positions at LIDS, MIT (USA) '12-'14, CERTH-ITI (Greece) '08-'12, and VTT (Finland), ‘07-‘08. For the period '09-'12 he also taught at the Electrical and Computer Engineering (ECE) Dept. of University of Thessaly. He received his diploma in ECE from Aristotle University of Thessaloniki, (’02) and his PhD degree in Wireless Networks from ECE dept. University of Patras (’06), both in Greece. Two of his papers won the best paper award, in GLOBECOM 07’ and IFIP Wireless Days 09’ respectively. He was the editor of IEEE JSAC special issue for content caching and delivery, and he actively serves as an associate editor for IEEE/ACM Trans. on Networking, IEEE Networking Letters, and as a Technical Program Committee member of INFOCOM, WiOPT, and Netsoft.
\end{IEEEbiography}

\begin{IEEEbiography}[{\includegraphics[width=1.25in,height=1.25in,clip,keepaspectratio]{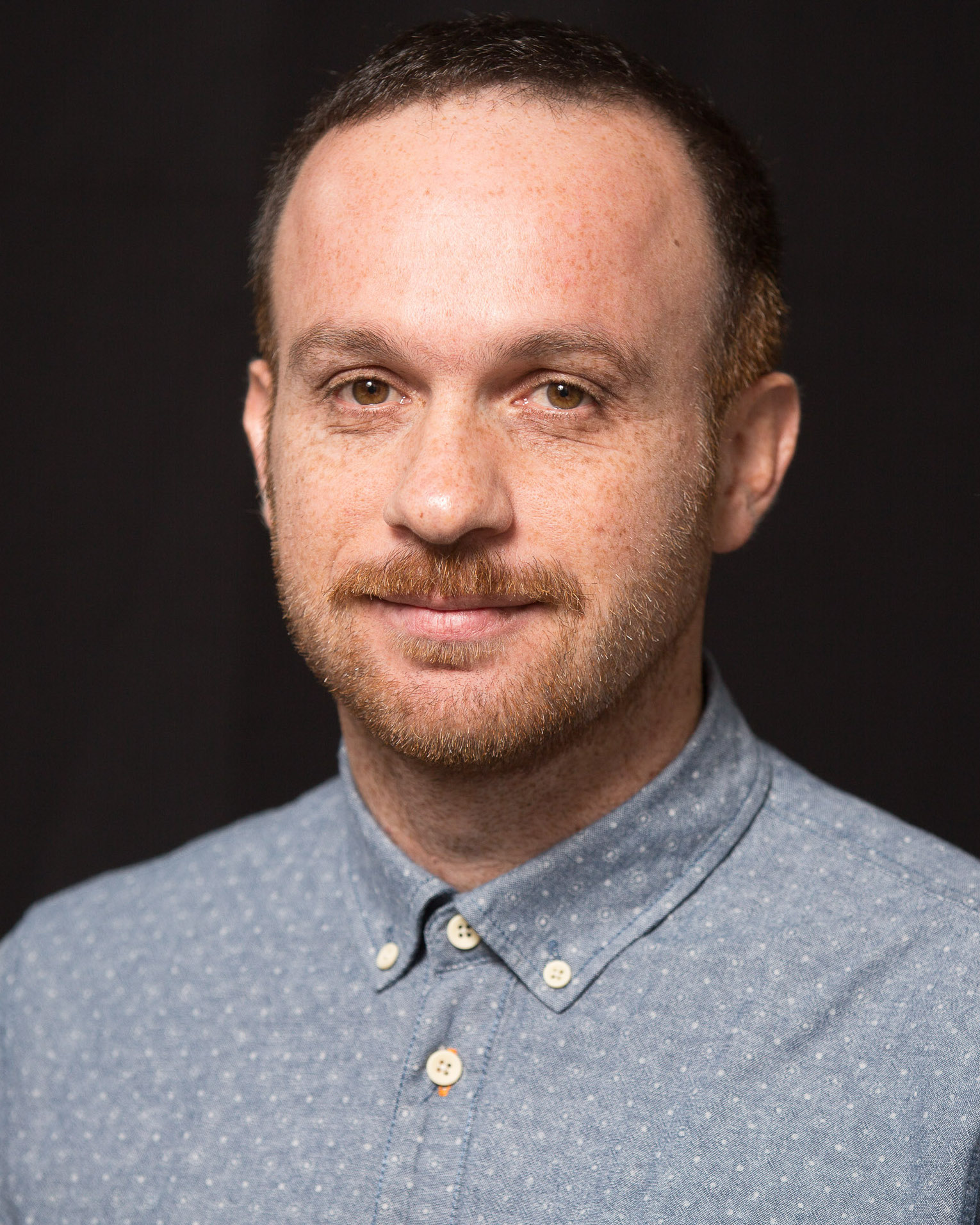}}] 
{George Iosifidis} is the Ussher Assistant Professor in Future Networks, at Trinity College Dublin, Ireland. He received a Diploma in Electronics and Communications, from the Greek Air Force Academy (Athens, 2000) and a PhD degree from the Department of Electrical and Computer Engineering, University of Thessaly in 2012. He was a Postdoctoral researcher ('12-'14) at CERTH-ITI in Greece, and Postdoctoral/Associate research scientist at Yale University ('14-'17). He is a co-recipient of the best paper awards in WiOPT 2013 and IEEE INFOCOM 2018 conferences, and has received an SFI Career Development Award in 2018.
\end{IEEEbiography}

\begin{IEEEbiography}[{\includegraphics[width=1.1in,height=2.5in,clip,keepaspectratio]{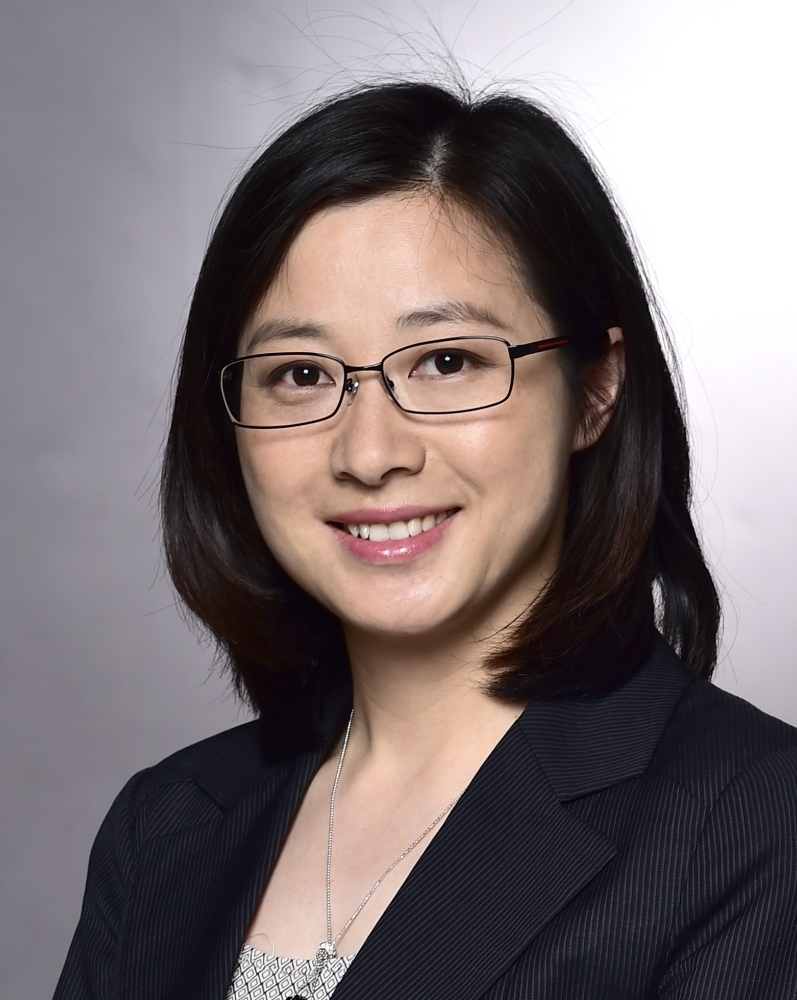}}]
{Meixia Tao} (S'00-M'04-SM'10) received the B.S. degree in electronic engineering from Fudan University, Shanghai, China, in 1999, and the Ph.D. degree in electrical and electronic engineering from Hong Kong University of Science and Technology in 2003. She is currently a Professor with the Department of Electronic Engineering, Shanghai Jiao Tong University, China. Prior to that, she was a Member of Professional Staff at Hong Kong Applied Science and Technology Research Institute during 2003-2004, and a Teaching Fellow then an Assistant Professor at the Department of Electrical and Computer Engineering, National University of Singapore from 2004 to 2007. Her current research interests include wireless caching, physical layer multicasting, resource allocation, and interference management.

Dr. Tao currently serves as a member of the Executive Editorial Committee of the \textsc{IEEE Transactions on Wireless Communications} and an Editor for the \textsc{IEEE Transactions on Communications}. She was on the Editorial Board of the \textsc{IEEE Transactions on Wireless Communications} (2007-20110), the \textsc{IEEE Communications Letters} (2009-2012), and the \textsc{IEEE Wireless Communications Letters} (2011-2015). She serves as Symposium Co-Chair of IEEE GLOBECOM 2018, the TPC chair of IEEE/CIC ICCC 2014 and Symposium Co-Chair of IEEE ICC 2015.

Dr. Tao is the recipient of the IEEE Heinrich Hertz Award for Best Communications Letters in 2013, the IEEE/CIC International Conference on Communications in China (ICCC) Best Paper Award in 2015, and the International Conference on Wireless Communications and Signal Processing (WCSP) Best Paper Award in 2012. She also receives the IEEE ComSoc Asia-Pacific Outstanding Young Researcher award in 2009.

\end{IEEEbiography}

\begin{IEEEbiography}[{\includegraphics[width=1.2in,height=1.2in,clip,keepaspectratio]{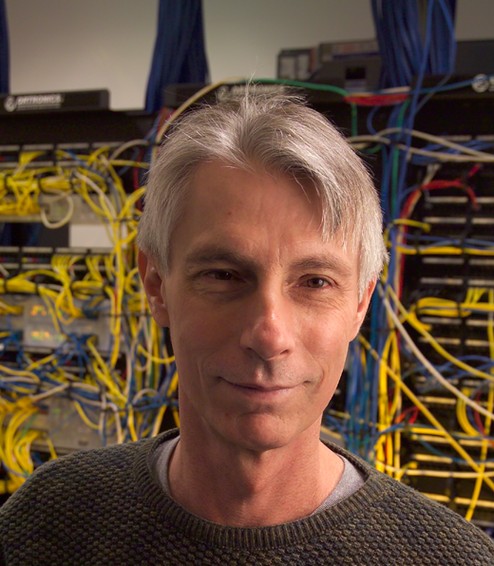}}]
{Don Towsley} holds a B.A. in Physics (1971) and a Ph.D. in Computer Science (1975) from University of Texas. He is currently a Distinguished Professor at the University of Massachusetts in the College of Information \& Computer Sciences. He has held visiting positions at numerous universities and research labs including University of Paris VI, IBM Research, AT\&T Research, Microsoft Research, and INRIA. His research interests include security, quantum communications, networks and performance evaluation. He is a co-founder ACM Transactions on Modeling and Performance Evaluation of Computing Systems (ToMPECS) and served as one of its first co-Editor in Chiefs. He served as Editor-in-Chief of the IEEE/ACM Transactions on Networking and on numerous other editorial boards. He has served as Program Co-chair for numerous conferences and on the program committees of many other. He is a corresponding member of the Brazilian Academy of Sciences and has received numerous IEEE and ACM awards including the 2007 IEEE Koji Kobayashi Award, 2007 ACM SIGMETRICS Achievement Award, and 2008 ACM SIGCOMM Achievement Award. He has also received numerous best paper awards including the IEEE Communications Society 1998 William Bennett Paper Award, a 2008 ACM SIGCOMM Test of Time Award, the 10+ Year 2010 DASFAA Best Paper Award, the 2012 ACM SIGMETRICS Test of Time Award and five ACM SIGMETRICS Best paper awards. Last, he is Fellow of both the ACM and IEEE.
\end{IEEEbiography}

\begin{IEEEbiography}[{\includegraphics[width=1.2in,height=1.2in,clip,keepaspectratio]{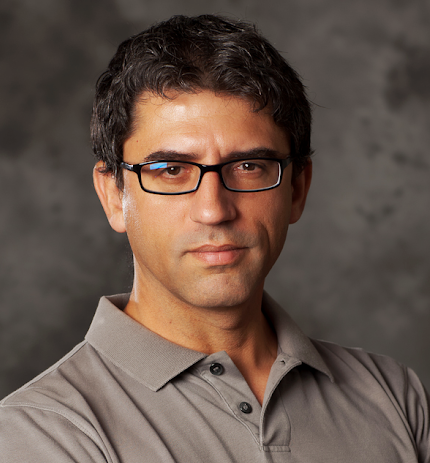}}]
{Giuseppe Caire} (S'92 - M'94 - SM'03 - F'05) was born in Torino, Italy, in 1965. He received the B.Sc. in Electrical Engineering from Politecnico di Torino (Italy), in 1990, the M.Sc. in Electrical Engineering from Princeton University in 1992 and the Ph.D. from Politecnico di Torino in 1994. He has been a post-doctoral research fellow with the European Space Agency (ESTEC, Noordwijk, The Netherlands) in 1994-1995, Assistant Professor in Telecommunications at the Politecnico di Torino, Associate Professor at the University of Parma, Italy, Professor with the Department of Mobile Communications at the Eurecom Institute, Sophia-Antipolis, France, and he is currently a professor of Electrical Engineering with the Viterbi School of Engineering, University of Southern California, Los Angeles and an Alexander von Humboldt Professor with the Electrical Engineering and Computer Science Department of the Technical University of Berlin, Germany. He served as Associate Editor for the IEEE Transactions on Communications in 1998-2001 and as Associate Editor for the IEEE Transactions on Information Theory in 2001-2003. He received the Jack Neubauer Best System Paper Award from the IEEE Vehicular Technology Society in 2003, the IEEE Communications Society \& Information Theory Society Joint Paper Award in 2004 and in 2011, the Okawa Research Award in 2006, the Alexander von Humboldt Professorship in 2014, and the Vodafone Innovation Prize in 2015. Giuseppe Caire is a Fellow of IEEE since 2005. He has served in the Board of Governors of the IEEE Information Theory Society from 2004 to 2007, and as officer from 2008 to 2013. He was President of the IEEE Information Theory Society in 2011. His main research interests are in the field of communications theory, information theory, channel and source coding with particular focus on wireless communications.
\end{IEEEbiography}

\end{document}

%% file: Section-IntroVer12.tex
\section{Introduction} \label{section:intro}

Today storage resources and caching techniques permeate almost every area of network and communication technologies. From storage-assisted future Internet architectures and  information-centric networks, to caching-enabled 5G wireless systems, caching promises to benefit both the network infrastructure (reducing costs) and the end-users (improving services). In light of pressing data traffic growth, and the increasing number of services that nowadays rely on timely delivery of (rich-media) content, the following questions are inevitably raised: \emph{can caching deliver on these promises?} and if the answer is affirmative, \emph{what are the required research advances to this end?} 
 
In this tutorial paper we investigate these two questions in detail. We start with a brief discussion about the historical background of caching, and then present three key factors that, in our opinion, render caching research very important today. These factors relate to constantly evolving user needs, novel demanding network services, but also to new technologies that can make caching very effective. In Section \ref{section:overview} we present the most active research areas in caching. We analyze seminal papers and discuss the latest developments in each area, and present the advances made by the papers appearing in this Special Issue. Our goal is to provide a unified view on the different (and often disconnected) research threads in caching.

In Section \ref{section:open-issues} we discuss several state-of-the-art caching systems, focusing on the research challenges they bring. We also present the latest wireless caching standardization efforts that pave the way for the design of new caching architectures. Finally we analyze a set of key open challenges, i.e., bottleneck issues that need to be resolved in order  to unleash the full potential of this promising tool. These issues range from the need to analyze the economic interactions in the complex caching ecosystem, to develop methods for coping with volatile content popularity, and to devise joint caching and computing solutions.

\subsection{Historical Perspective}

The term \emph{cache} was introduced in computer systems to describe a memory with very fast access but typically small capacity. By  exploiting correlations in memory access patterns, a small cache can significantly improve system performance. Several important results related to online caching strategies can be found in papers from the 1970s. Prominent examples include the oracle MIN policy that maximizes hits under an arbitrary request sequence \cite{belady}, and the analysis of Least-Recently-Used (LRU) policy under stationary sequences using a Markovian model \cite{King71} or using an efficient approximation \cite{fagin}.

The caching idea was later applied to the Internet: instead of retrieving a webpage from a central server, popular webpages were replicated in smaller servers (\emph{caches}) around the world, reducing \emph{(i)} network bandwidth usage, \emph{(ii)} content access time, and \emph{(iii)} server congestion. With the rapid Internet traffic growth in late 1990s, the management of theses caches became complicated. This led to the proliferation of \emph{Content Delivery Networks} (CDNs), an integral part of the Internet ecosystem that employ monitoring and control techniques to manage interconnected caches. Research in CDNs made progress in the last decades on investigating \emph{(i)} where to deploy the servers (\emph{server placement}) \cite{qiu-infocom01}, \emph{(ii)} how much storage capacity to allocate to each server (\emph{cache dimensioning}) \cite{Kelly01}, \emph{(iii)} which files to cache at each server (\emph{content placement}), and \emph{(iv)} how to route content from caches to end-users (\emph{routing policy}). However, new questions arise today as CDNs need to support services with more stringent requirements.

Recently, caching has also been considered for improving content delivery in wireless networks \cite{misconceptions}. Indeed, network capacity enhancement through the increase of physical layer access rate or the deployment of additional base stations is a costly approach, and outpaced by the fast-increasing mobile data traffic \cite{cisco, ericsson}. Caching techniques promise to fill this gap, and several interesting ideas have been suggested: \emph{(i)} deep caching at the {evolved packet core} (EPC) in order to reduce content delivery delay \cite{caching-cellular}; \emph{(ii)} caching at the base stations to alleviate congestion in their throughput-limited backhaul links \cite{Dimakis:femtocaching}; \emph{(iii)} caching at the mobile devices to leverage device-to-device communications \cite{caired2d}; and \emph{(iv)} coded caching for accelerating transmissions over a broadcast medium \cite{nielsen}. There are many open questions in this area, and several papers of this Special Issue focus on this topic.

\subsection{Caching for Future Networks}

There is growing consensus that caching is poised to play a central role in future communication systems and networks, and inevitably the following question arises: \emph{Can we tackle the upcoming challenges in caching using existing tools?} We believe that the answer to this question is an emphatic ``no'', providing motivation to further study caching systems. Our belief is based on three main arguments that we summarize in the remainder of this article. 

\emph{The content demand characteristics} continue to evolve. Internet-based online video gradually replaces classical Television, and new specifications (4K, QHD, 360$^o$, etc.) increase the bandwidth consumption per content request. Furthermore most video files need to be available in different encoding format, and this \emph{versioning} enlarges caching requirements. These factors drive the explosion of video traffic, which is expected to surpass $80\%$ of the total Internet traffic \cite{cisco}. At the same time, new services are emerging, such as (mobile) Augmented and Virtual Reality with even tighter bandwidth and latency requirements than typical video streaming. 
{
These services will constantly feed the users with enormous amounts of personalized sensory information and hologram depictions in real time, and hence have to rely on local edge caches.} 
Finally, the proliferation of online social networks (OSNs) is placing users in the role of content creator, and disrupting the traditional server-user client model. OSNs increase the volatility of content popularity, and create often unforeseen spatio-temporal traffic spikes. In sum, the characteristics of \emph{cache-able} content and content demand are rapidly changing, forcing us to revisit caching architectures and caching solutions. 

\emph{Memory as a fundamental resource}. Recent developed techniques that combine caching with coding demonstrate revolutionary \emph{goodput} scaling in bandwidth-limited cache-aided networks \cite{nielsen}. This motivated the community to revisit the fundamental question of how memory ``interacts'' with other types of resources. Indeed, the topic of \emph{coded caching} started as a powerful tool for broadcast mediums, and is being currently expanded towards establishing an information theory for memory. The first results provide promising evidence that the throughput limits of cache-enabled communication systems are in fact way beyond what is achievable by current networks. Similarly, an interesting connection between memory and processing has been recently identified \cite{cdc}, creating novel opportunities for improving the performance of distributed and parallel computing systems. These lines of research have re-stirred the interest in joint consideration of bandwidth, processing, and memory, and promise novel caching systems with high performance gains.

\emph{Memory cloudification and new architectures}. Finally, the advent of technologies such as Software-Defined Networking (SDN) and Network Function Virtualization (NFV) create new opportunities for leveraging caching. Namely, they enable the fine-grained and unified control of storage capacity, computing power and network bandwidth, and facilitate the deployment of in-network caching services. Besides, recent proposals for content-centric network architectures place storage and caching at a conspicuous place, but require a clean-slate design approach of caching techniques. At the same time, new business models are emerging today since new players are entering the content delivery market. Service providers like Facebook are acquiring their own CDNs, and network operators deploy in-network cache servers to reduce their bandwidth expenditures. These new models create, unavoidably, new research questions for caching architectures and the caching economic ecosystem.

\subsection{About this Issue}

\begin{table}
	\centering
	\caption{Caching Topics Statistics} 
	\begin{tabular}{| l | c | c | }
		\hline
		\textbf{Topic and Keywords}   		& \textbf{Papers}  \\ \hline
		Information-theoretic Caching Analysis      & 16       \\ \hline
		Fundamental Limits of Caching   			& 16      \\ \hline
		Coded Caching Design with Practical Constraints    	& 12       \\ \hline
		Scaling Laws of Cache Networks						& 7       \\ \hline
		HetNet and Device-to-device Caching		&  18      \\ \hline
		Edge Caching, Cooperation and Femtocaching		&  32     \\ \hline
		Joint Caching, Scheduling and Routing		& 19      \\ \hline
		Secure Caching and Privacy Preservation		& 6      \\ \hline
		Content Caching and Delivery		&  56      \\ \hline
		Algorithms for Storage Placement		&  17     \\ \hline
		Video caching and Streaming		&  14      \\ \hline
		Caching Economics		&   13    \\ \hline
		Caching models for ICN		&  12     \\ \hline
		Popularity Models and Machine Learning		& 9      \\ \hline				
	\end{tabular}
	
	\label{table:topic-statistics}
\end{table}

This Special Issue received a very large number of submissions verifying that caching is an active research topic in many areas: 237 authors from Asia/Pacific ($50.6\%$ of total), 121 from Europe, Middle East, Africa ($25.9\%$), 105 from the United States and Canada ($23.5\%$). These statistics show that caching appears to be most popular in P.R. China, USA, Korea, UK and France. During the review process approximately 360 experts were involved, and this indicates the large body of researchers on this topic. Table \ref{table:topic-statistics} shows a breakdown of topics in the submissions where up to three were registered per paper from its authors. These numbers are indicative of the current popularity of each topic.

The final version of the ``JSAC-caching'' Special Issue comprises novel technical contributions aiming to address a wide span of caching challenges for communication systems and networks. The topics include \emph{information theory and coded caching}, \emph{caching networks}, \emph{caching policies and storage control}, \emph{wireless caching techniques}, \emph{caching economics}, and \emph{content-based architectures}. In the following section we visit each research direction in detail, explaining the main idea and discussing the new contributions made in this Special Issue.

%% file: Section-Overview-1Ver12.tex
\section{Caching: Past and Present}\label{section:overview}

This Section presents the background, seminal papers, and recent developments in important research areas of caching. Furthermore, we present the papers that appear in this Issue and explain how they advance the state-of-the-art literature.

\begin{figure}[t]
	\centering
	\includegraphics[width=0.94\linewidth ]{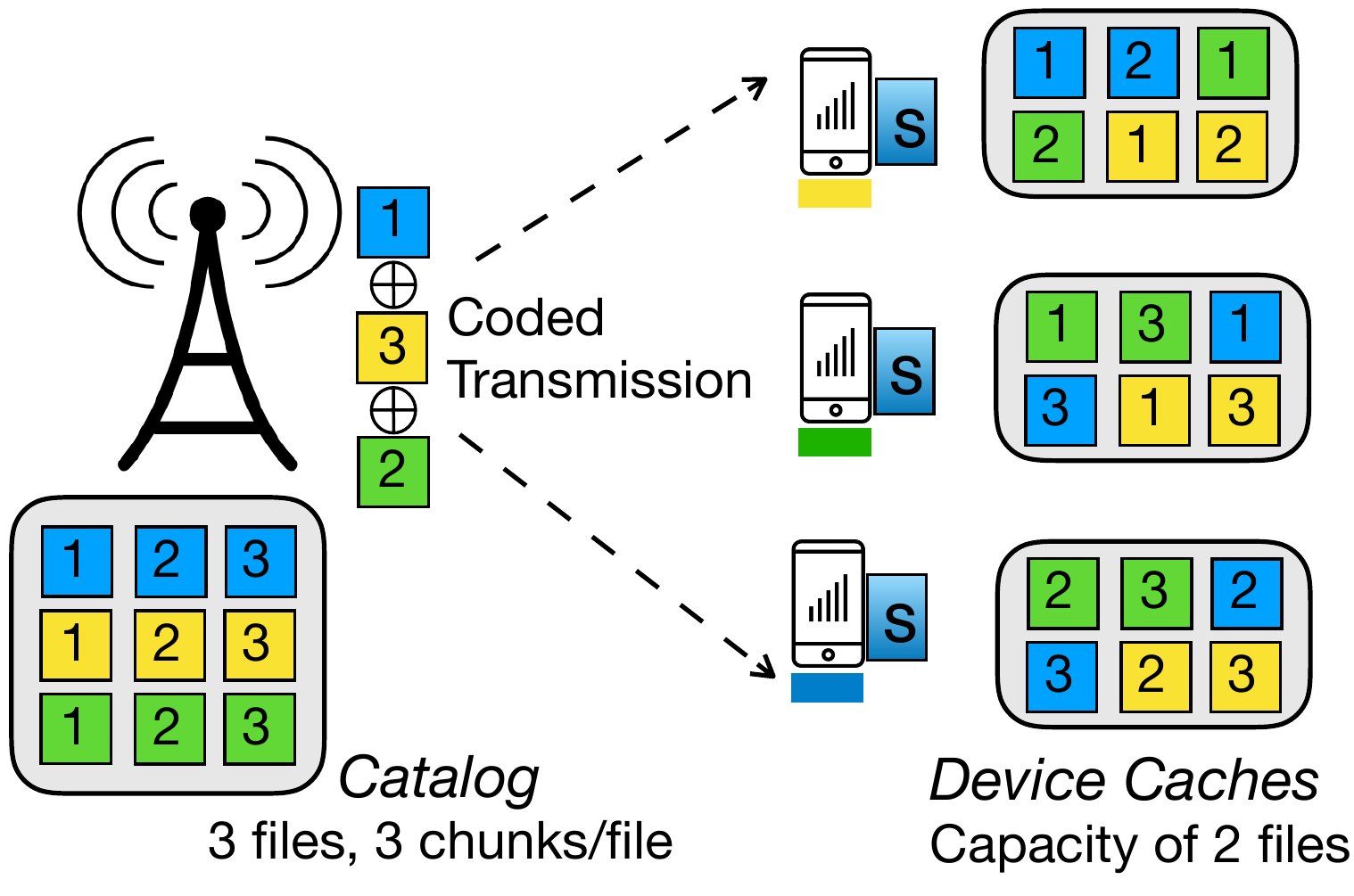}   	
	\caption{\small Coding increases the cache performance \cite{nielsen}.}
	\label{fig:coded-caching}    
\end{figure}

\subsection{Information-theoretic Caching Analysis}

In 2013 Maddah-Ali and Nielsen studied the fundamental limits of broadcast transmissions in the presence of receiver caching \cite{nielsen}. The problem was defined as follows: assume a shared error-free medium connecting a source to $K$ users, each one requesting one file of $F$ bits. There are $N$ available files in the system (in total $NF$ bits) while each receiver can store $MF$ bits in their cache, with $\frac{M}N\triangleq \gamma<1$ denoting the relative cache size. A caching policy $\pi$ performs two functions: \emph{(i)} placement during which it decides the bits (or functions of bits) that will be stored at each cache spot, and \emph{(ii)} delivery during which it determines a sequence of multicast transmissions that ensures correct delivery, i.e. that every receiver  obtains the requested file. We denote by $R$ the vector of file requests, and by $T^{\pi}(R)$ the smallest number of transmissions under policy $\pi$ such that all receivers have obtained their files indicated by $R$. The problem is to find $\pi$ that attains $\inf_{\pi}\max_{R}T^{\pi}(R)$. Note that traditional caching policies would place a $\gamma$ fraction of all files, requiring $K(1-\gamma)$ transmissions under any request.

\begin{figure}[t!]
	\centering
		\includegraphics[width=0.996\linewidth ]{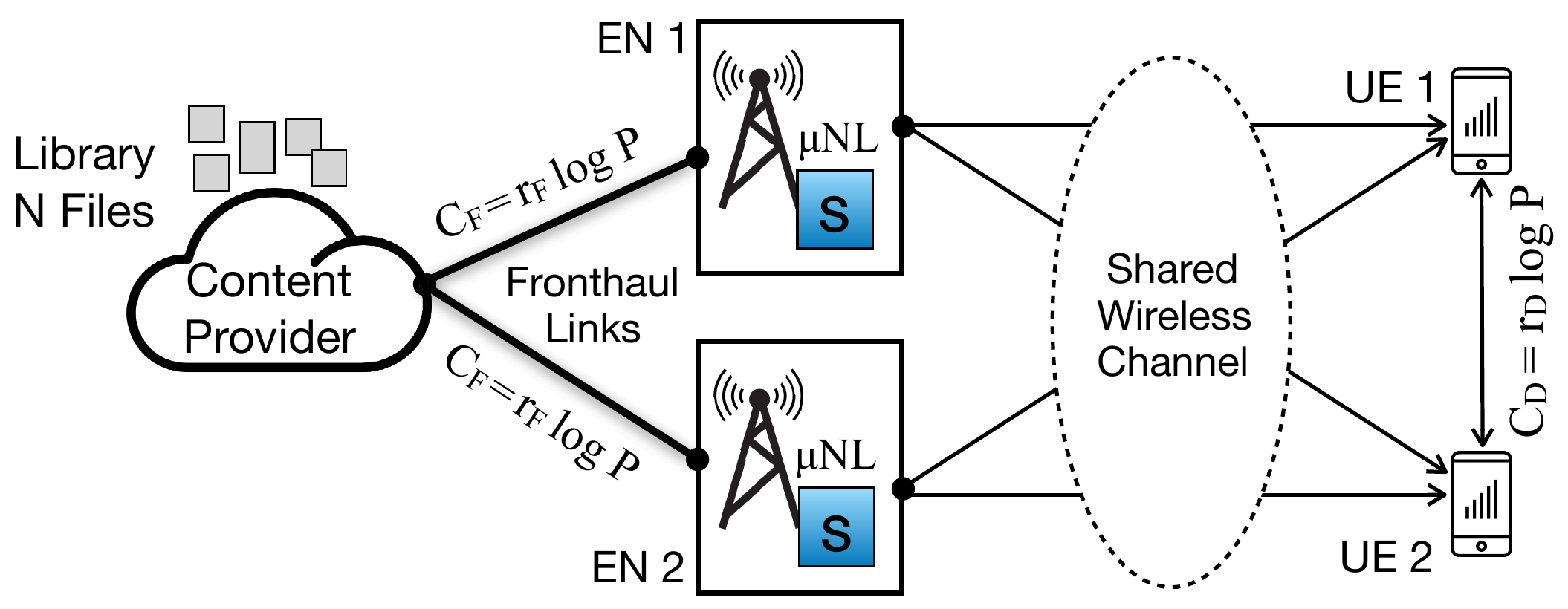}   	
	\caption{\small  Minimizing latency with Tx coded caching in a system with cache-enabled transmitters \cite{simeone-d2d-aided-2018}.
}
	\label{fig:c-ran}    
\end{figure}

Although this problem is largely intractable, the seminal paper  \cite{nielsen} proposed a scheme that achieves a 12-approximation. The policy known as ``centralized coded caching'' is depicted in Fig. \ref{fig:coded-caching}. During placement the policy splits the caches in parts corresponding to all subsets of users and caches different  bit combinations. During delivery, the existence of bit XOR combinations is guaranteed such that the scheme is correct with at most $K(1-\gamma)/(1+\gamma K)$ transmissions. This provides a $1+\gamma K$ gain over classical caching. Further, the number of required transmissions converges to $\frac{1-\gamma}{\gamma}$ for $K\to \infty$. This implies that in a wireless downlink with finite resource blocks, an indefinite number of memory-equipped receivers can be simultaneously served (albeit at a small rate). Several extensions of this scheme were subsequently considered, e.g., the scenario of adding storage capacity at the transmitters aiming to reduce latency \cite{simeone-d2d-aided-2018}, Fig. \ref{fig:c-ran}.

A large number of papers appearing in this Special Issue are related to coded caching. The classical coded caching scheme suffers from the subpacketization issue; maximum gains  can be achieved only if the packets are split into $2^K$  pieces. Since an $L$-bit packet can be split at most $L$ times (typically much less in practical systems), as $K$ increases the gains diminish. This problem is studied in \cite{coded-lampiris} which suggests the addition of antennas to the source. In particular it shows that $W$ transmit antennas reduces the required subpacketization to approximately its $W$-th root. Similarly, \cite{coded-malik} studies the throughput-delay trade-offs in an ad hoc network where each node moves according to a simplified \emph{reshuffling mobility model}, and extends prior work to the case of subpacketization.

Reference \cite{coded-zhang} introduces a novel unification of two extreme and different approaches in coded caching, namely \emph{(i)} the uncoded prefetching designed by \cite{nielsen}, and the \emph{(ii)} the coded prefetching designed in \cite{Tian16}. A scheme that  generalizes both prior cases is proposed, and it is shown that it achieves new trade-offs. On the other hand \cite{coded-gomez} uses coded prefetching to achieve the rate-memory region of \cite{Tian16} with a much smaller field (of order $2^2$ instead of $2^m, m\geq K\log_2 N$).

Several recent works focus on aspects related to user demand. In \cite{coded-poor} the coded caching framework is extended to an online approach that designs placement and delivery while considering user demand asynchronism. In the same context,  \cite{coded-azimi} extends the information theoretic analysis of caching to the case of C-RAN clouds with time-varying popularity, and looks at the normalized delivery time metric which captures both the delivery part of coded caching and the time needed to replenish the cache contents. In \cite{coded-tulino} the idea of coded caching is generalized to correlated content showing how to exploit correlations to obtain a lower bound for the rate-memory trade-off. The paper \cite{coded-hachem} matches users to one of a subset of caches after the user request is revealed. It compares a scheme focusing on coded server transmissions while ignoring matching capabilities, with a scheme that focuses on adaptive matching while ignoring potential coding opportunities.


The paper \cite{coded-amiri} extends coded caching to wireless broadcast channels with Gaussian noise, and studies the optimization of energy efficiency. Reference \cite{coded-mahdian} addresses the problem of combining network coding and caching in order to optimize a multicast session over a directed-acyclic graph. In \cite{coded-wei} the authors study the fundamental limits of secretive coded caching, examining the case of partially known prefetched files. The work in \cite{coded-zewail} also considers secrecy constraints; it studies a two-hop network architecture known as a combination network, where a layer of relay nodes connects a server to a set of end users. A new centralized coded caching scheme is developed that jointly optimizes cache placement and delivery phase, and enables decomposing the combination network into a set virtual multicast sub-networks.

\subsection{Caching in Wireless Systems}

\begin{figure*}
	\centering
	\subfigure[Core-caching]{
		\includegraphics[width=0.35\linewidth ]{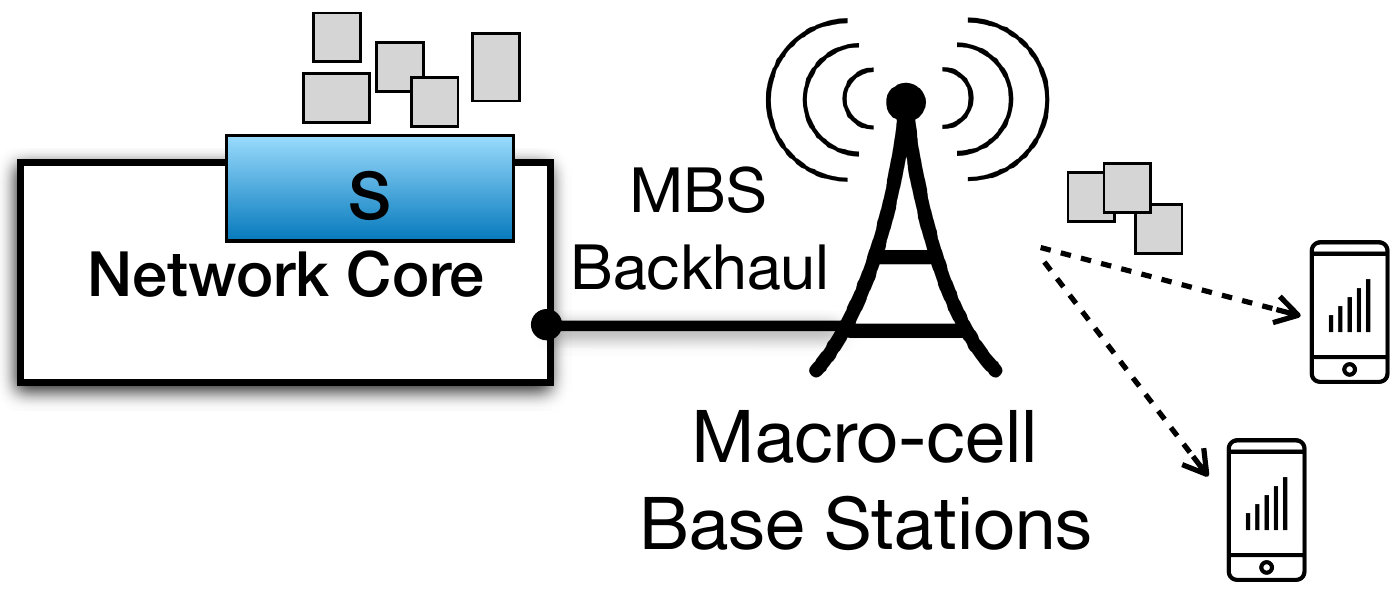}
		\label{fig:wireless-core}}
	\subfigure[Femtocaching]{
		\centering
		\includegraphics[width=0.25\linewidth]{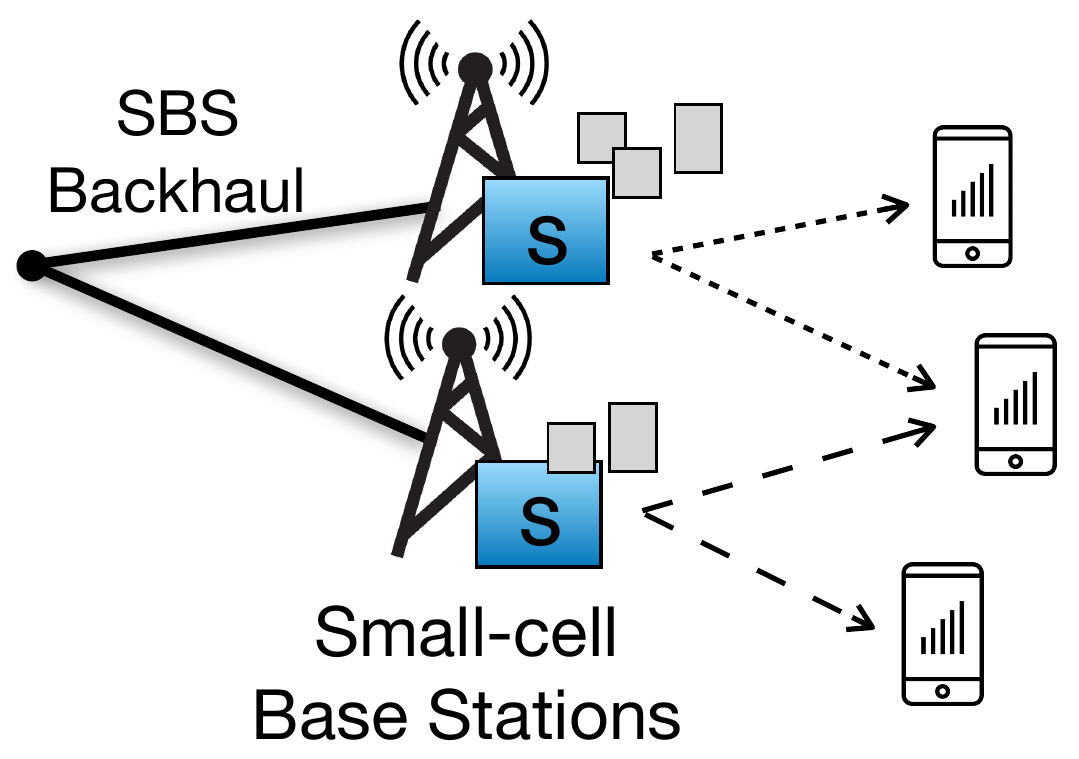}
		\label{fig:femtocaching}}
	\subfigure[D2D Caching]{
		\centering
		\includegraphics[width=0.28\linewidth]{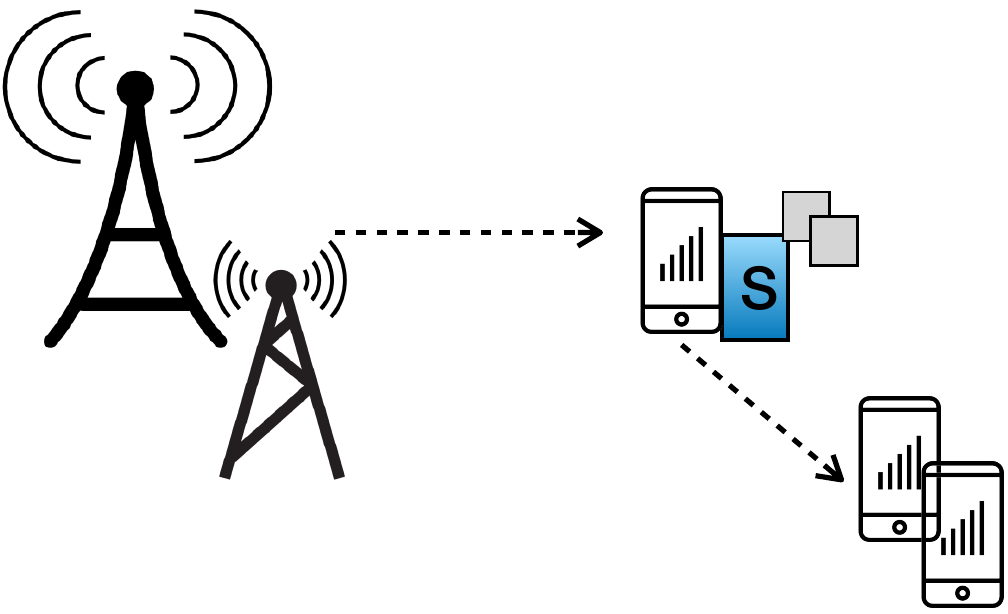}
		\label{fig:topo:turin}}    	
	\caption{\small Different scenarios for placing storage (``S'') at wireless networks. (a): Caching at the evolved packet core of a mobile network; (b): Caching at small cell base stations (Femtocaching); (c): Caching at the user device and device-to-device (D2D) content delivery.}
	\label{fig:wireless-caching}    
\end{figure*}

Beyond coded caching there are several recent proposals for novel cache-aided wireless network architectures, and for techniques that combine caching with other wireless communication decisions.

\subsubsection{Femtocaching and D2D} Caching content at the very edge of a wireless networks (base station; user devices) is fundamentally different from caching techniques in CDN systems and raises novel challenges. Namely, in wireless networks the demand per edge cache is smaller in volume which varies rapidly with time as users move from one cell to another. Furthermore, caching decisions are coupled not only because caches share backhaul links, but also because users might be in range of multiple cache-enabled base stations. These characteristics, together with the inherent volatility of the wireless medium, render caching decisions particularly difficult to optimize and, oftentimes, less effective e.g., in terms of the achieved cache hit ratio.

Nevertheless, several interesting proposals for wireless caching have recently appeared, Fig. \ref{fig:wireless-caching}. The seminal ``femtocaching'' paper \cite{Dimakis:femtocaching} proposed the idea of proactive caching at small cell base stations as a solution to their capacity-limited backhaul links. The problem of minimizing the average content delivery delay was formulated and solved using  submodular optimization. Many follow-up works focus on this architecture, including \cite{wireless-shukla} in this Issue (discussed later). In a similar setting, \cite{[Ji-D2D-15]} studied content dissemination through device-to-device (D2D) communications. It was shown that short-range D2D transmissions combined with content caching at user devices yield a scaling law of throughput which is independent of the number of users.

\subsubsection{Caching and Wireless Transmissions}

The design of wireless transmission techniques changes significantly in the presence of caching. For example, caching at transmitters can turn an interference channel into a broadcast channel or X-channel \cite{[MN-isit15]}; and caching at both transmitters and receivers can turn an interference channel into a so-called cooperative X-multicast channel \cite{[Xu-TIT17]}. Clearly, physical-layer transmission and scheduling schemes have to be re-visited in cache-enabled wireless networks. 

The new cache-aided designs induce a coupling between the transmissions and the caching strategies, and this gives rise to challenging mixed time-scale optimization problems. For example, \cite{[Liu-TSP13]} showed that by caching a portion of each file, the base stations can opportunistically employ cooperative multipoint (CoMP) transmission without expensive backhaul in MIMO interference networks, yielding the so-called cache-induced opportunistic CoMP gain; a technique that requires the joint optimization of MIMO precoding (small time scale) and caching policies (large time scale). The joint design of caching, base station clustering, and multicast beamforming can significantly improve the energy-backhaul trade-offs in C-RAN systems \cite{[Tao-TWC16]}. More complicated cross-time-scale interactions are investigated in \cite{[Abedini-TON14],[Poularakis-TWC16],[Zhou-TWC16]} for either throughput maximization or service cost minimization.

\subsubsection{Caching in Stochastic Wireless Networks}

Another line of research that has attracted great attention is the caching optimization in stochastic wireless networks where node locations are modeled as independent spatial random processes, e.g., Poisson Point Process (PPP). Due to advances in stochastic geometry tools for cellular networks, cf. \cite{[Andrews-TCOM11]}, this approach facilitates the analysis and design of large-scale wireless caching systems. Assuming  that base stations cache the most popular contents, the works \cite{[Bastug-WCN15]}, \cite{[Yang-TWC16]} derived closed-form expressions for the outage probability and the average delivery rate of a typical user as a function of SINR, base station density, and file popularity. If base stations cache the contents randomly, the optimization of  caching probabilities is considered in \cite{[Chen-TVT17]}. If base stations employ maximum distance separable (MDS) codes or random linear network codes for content caching, the optimization of caching parameters is considered in \cite{[Xu-TCOM17]}.




Reference \cite{wireless-xu} extends caching to communication scenarios with Unmanned Aerial Vehicles (UAVs). Specifically, it proposes policies that decide jointly caching and trajectories of UAVs in order to maximize the efficiency of content delivery. The authors in \cite{wireless-zhang} investigate video caching over heterogeneous networks modeled with PPP, and study the impact of different viewing quality requirements on energy efficiency. \cite{coded-deng} uses stochastic geometry to model the locations of base stations and study different cooperative caching techniques, including coded caching. The proposed solutions demonstrate superior energy efficiency for the network over selected benchmark schemes. The idea of improving caching decisions in small cells and user devices by considering social-layer characteristics, such as mutual user interests and mobility patterns, is proposed in \cite{edge-caching-li}. Finally, a mixed time-scale problem is studied in \cite{dai-jsac18} with cache dimensioning at the base stations and beamforming decisions for improving content delivery in C-RAN systems.


\subsection{ICN Architectures}

Information (or, Content) Centric Networking (ICN or CCN) is a research thread that aims to provide a clean-slate design of the Internet \cite{jacobson09}. The main idea is to redesign basic communication functions (such as routing) based on content addressing, replacing the IP-based network paradigm, cf. this survey \cite{xylomenos-survey}. As the traffic volume of video and other types of content grow fast, ICN architectures are becoming increasingly relevant. 

The core idea in ICN is to proactively publish content to all interested Internet entities using a multicast session. This type of  communications are inherently related to caching and motivate  novel technical questions: \emph{(i)} how to perform en-route caching with mechanisms such as ``leave a copy'' \cite{sourlas_enroute}; \emph{(ii)} how to design caching structures that can scale to handle large traffic volumes; and \emph{(iii)} how much storage to deploy in the network, and at which nodes, so as to balance costs and performance gains \cite{roberts}. 

Another crucial topic in ICN is content discovery. While collaboration of caches increases the hit performance (by fine-tuning how often each file is replicated), in ICN this promising architecture entangles the content discovery process. Namely, in a network of caches, although a replica can be cached closer to the user, discovering its actual location may take a significant amount of time and even violate the QoS criteria due to excessive delays. 

The problem of content discovery is studied in \cite{icn-crowcroft}, which proposes the \emph{scope-flooding} technique to propagate control signals for content discovery by building multicast trees routed at the source node. Since replica frequency is expected to relate to popularity, the authors suggest tuning the discovery radius according to content popularity.

Another important direction for ICN is certainly the efficient simulation of large caching installations. Due to the immense number of contents, it might be computationally-demanding (and even prohibitive) to model and simulate such systems, e.g., in order to assess the performance of different caching policies. The work of \cite{simulations-tortelli} revisits this problem and proposes model-driven techniques for simulating general cache networks. This solution leverages very accurate approximations of caching systems to be able to allow the simulation of hundreds of caches and trillions of contents, while employing complex routing and caching algorithms.

\subsection{Online Caching Policies and Analytics}

\begin{figure}[t!]
	\centering
	\includegraphics[width=0.996\linewidth ]{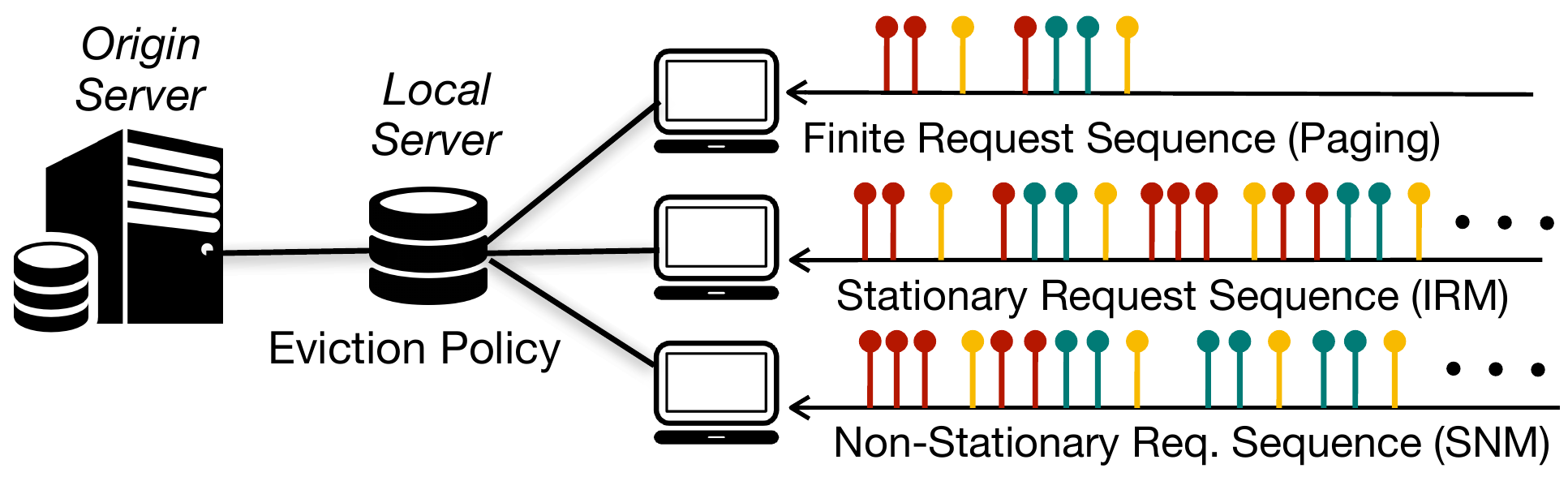}   	
	\caption{\small Examples of different eviction policies under different request arrival patterns.}
	\label{fig:eviction}    
\end{figure}

Online caching refers to the problem of devising a cache eviction policy in order to maximize the cache hit probability. This is a fundamental and well-studied topic, yet remains highly relevant today. The problem definition involves a cache of certain size, a file request sequence, and the eviction policy that determines which content should be removed when the cache overflows, Fig. \ref{fig:eviction}. Typical versions of this problem consider: \emph{(i)} finite request sequences and aim to devise eviction policies that maximize the number of hits; \emph{(ii)} stationary request sequences, with the goal to maximize the hit probability (stationary behavior); and \emph{(iii)} non-stationary sequences, where an additional challenge is to track the evolution of content popularity. 

Various eviction policies have been proposed in the past, each one having different advantages. For example, the Least-Recently-Used (LRU) policy promotes file recency and optimizes performance under adversarial finite request sequences (achieves the optimal competitive ratio \cite{sleator85}). Similarly, Least-Frequently-Used (LFU) policies maximize hit ratio under  stationary sequences by promoting the contents with the highest request frequency, while Time-To-Live (TTL) policies use timers to adjust the hit probability of each content \cite{massoulie-utility-cache}. Lately, great emphasis is put on online caching with non-stationary request sequences, focusing on the practical, yet challenging, scenario of time-varying file popularity.

Indeed, when popularity varies with time caching performance can not be determined solely based on the stationary hit ratio, and the eviction policy needs to be able to adapt to content popularity changes. In order to shed light on this aspect, \cite{learning-li} studies the mixing time of caching policies. It suggests that (most) eviction policies can be modeled as Markov chains, whose mixing times give us a figure of how ``reactive'' the  policy is, or else how true to its stationary performance. The $\tau$--distance \cite{kendal_tau} is leveraged for characterizing the learning error of caching policies. A practical lesson learned is that although multi-stage LRU policies offer tunable hit rate performance, they  adapt slowly to popularity changes.

Another line of research employs prediction schemes to accurately exploit file popularity instead of relying on LRU-type eviction rules. The authors of \cite{learning-gunduz} employ reinforcement learning in order to keep track of file popularity. They study proactive caching and content delivery in wireless networks and minimize the average energy cost. The model includes wireless links with time-varying channel quality, a time-evolving content catalog, and non-stationary file popularity. In \cite{wireless-zhao} traces from a vehicular ad-hoc network are used, and it is demonstrated that prediction-enhanced content prefetching can indeed increase the network performance. In this case the predictions refer both to content popularity and the vehicles (end-users) location. 

Reference \cite{learning-sermpezis} argues that users often have elastic needs and can be satisfied with similar content, if the requested items are not available at a local cache, which results in a  \emph{Soft Cache hit}. This work is in line with the recently proposed idea of leveraging recommendation systems that are embedded in several CDNs (e.g., YouTube) in order to steer user demand towards already cached content \cite{iordanis-caching}.

Finally, the problem of online cooperative caching (femtocaching) is considered in \cite{wireless-leonardi}, which is essentially a multi-cache generalization of the classical paging problem. The authors propose the ``lazy'' qLRU policy, where only the cache that is serving a content item can update its state, and it does so with probability $q$. It is shown that as $q\to 0$, the performance of this policy achieves a stationary limit which is a local maximum, in the sense that no performance improvement can be obtained by changing only one content's placement frequency.

%% file: Section-Overview-2Ver12.tex
\subsection{Content Caching and Delivery Techniques}

\begin{figure}
	\centering
		\includegraphics[width=0.95\linewidth ]{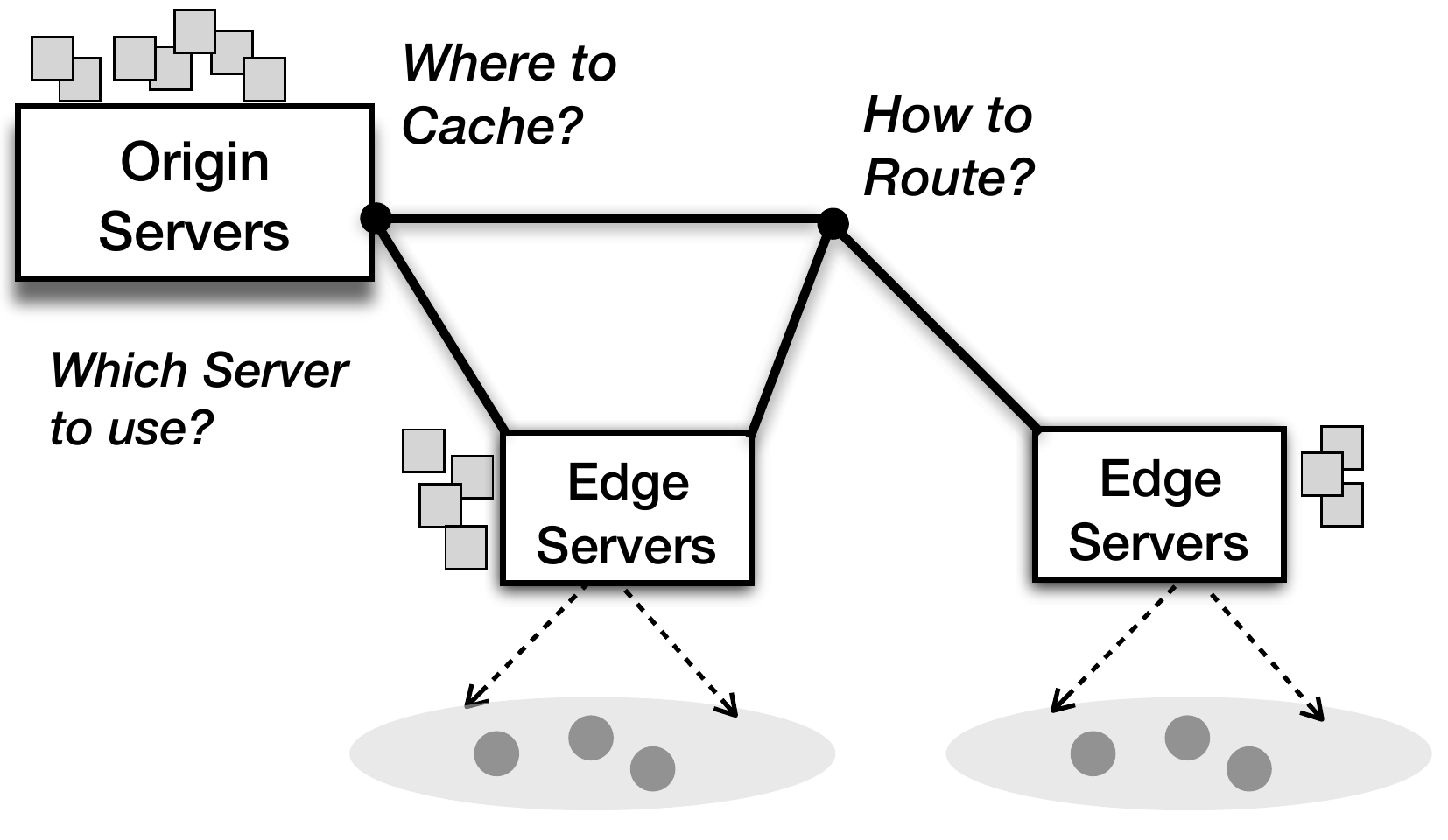}
	\caption{\small Decisions in Caching Networks \cite{bektas-1}. \textbf{Small time scale}: which server to use? where to cache? and how to route the content? The caching and routing decisions are inherently coupled, as a request can only be routed to cache where the requested item is available. \textbf{Large time scale}: where to place servers, and how to dimension the links connecting them?}
	\label{fig:hierarchical}    
\end{figure}


Modern caching systems are essentially networks of interconnected caches. Therefore, the caching problem in its entirety includes decisions about \emph{server placement} and \emph{cache  dimensioning}, \emph{content placement}, and \emph{content routing}, Fig. \ref{fig:hierarchical}. One can consider additional design parameters for these \emph{caching networks} (CNs) as, for example, dimensioning the links and the cache serving capacity.   

\subsubsection{Cache Deployment and Dimensioning} The storage (or, cache) deployment has been extensively studied and we refer the reader to \cite{sahoo-survey} for a survey. The cache deployment problem aiming to minimize content delivery delay has been formulated as a $K$-median problem \cite{qiu-infocom01}, \cite{krishnan-ToN00}, and a facility location problem \cite{bateni-TransAlgo12}. The work \cite{cronin-jsac02} studies a variation considering the cost of syncing different caches (ensuring consistent copies). It is shown that increasing the number of caches beyond a certain threshold induces costs higher than the performance benefits. When the network has a tree structure these deployment problems can be solved efficiently using dynamic programming \cite{benoit-TPDS08}, \cite{li-infocom99}.


\subsubsection{Hierarchical Caching Networks} 
\begin{figure}
	\centering
	\includegraphics[width=0.85\linewidth ]{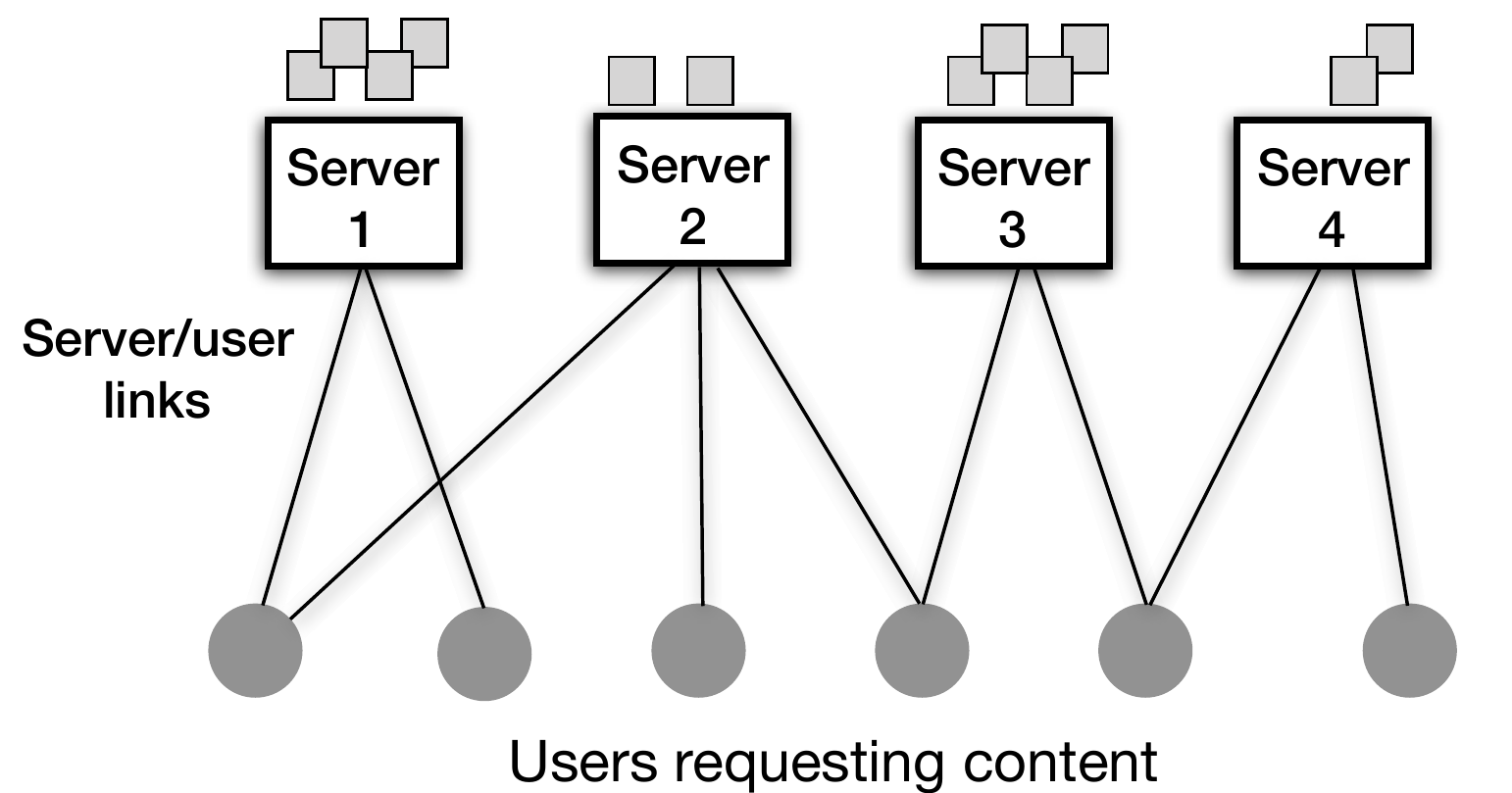}
	\caption{\small Bipartite caching model \cite{Dimakis:femtocaching}. A set of users is connected with caching servers. Every user can fetch content from each server with different cost, and servers cache possibly different items. The bipartite model is generic and captures several wired and wireless architectures, where link cost parameters can represent delay, energy, or monetary costs.}
	\label{fig:bipartite}    
\end{figure}

Indeed, CDNs or IPTV networks have often a tree-like form which facilitates the design of caching and routing policies. Caching at leaf nodes improves the access time, while caching at higher layers increases the cache hit ratio. Hierarchical networks are typically studied for 2-layers, often with the addition of a distant root server. The seminal work \cite{korupulu-hierarchical} presented a polynomial-time exact algorithm for the min-delay content placement problem when leaf caches can exchange their files. Motivated by an actual IPTV network, \cite{dai-collabor} studied a similar problem for a 3-level hierarchical caching network. Also \cite{pacifici-itc16} considers hierarchical caching for a wireless backhaul network and designed a distributed $2$-approximation caching algorithm. Another important objective is to minimize the requests sent to the root server in order to reduce off-network bandwidth and server congestion \cite{poularakis-TCOM16}. In many cases, these multi-tier CNs can be modeled also as bipartite caching networks Fig. \ref{fig:bipartite}, where the links capture the cost of the entire path connecting the user with each cache.

\subsubsection{General Caching Networks} There are also general CN models \cite{bektas-exact}, \cite{carofiglio-comnets16} where the link delay increases non-linearly with the number of requests \cite{towsley-Ton17}, or the objective functions are non-linear to the cache hit ratio \cite{poular-infocom14}. In some of these cases the problem has a convenient convex or submodular structure and hence greedy algorithms ensure a 2-approximation ratio \cite{towsley-Ton17}. Finally, in the general case, routing can involve both multihop and multipath decisions. This means that, in the presence of hard capacity constraints or load-dependent costs, the routing decisions are not fixed for a given content placement policy (as, e.g., in femtocaching) but need to be jointly devised.


The paper \cite{wireless-zhao} that appears in this Special Issue presents the interesting scenario of a 3-tier vehicle ad hoc network that includes origin servers, regional servers and road-side units. Focusing on general network architectures, \cite{caching-networks-ioannidis} studies the minimum-cost joint routing and caching problem and shows the benefit over considering the two problems separately. The study includes both hop-by-hop routing and source-routing decisions, and proposes distributed and online solution algorithms that achieve constant approximation ratio in polynomial time. Finally, in \cite{wireless-shukla} the femtocaching problem is being extended in the setting where files can be stored for a limited duration at a cache and delivered with multicast. The authors provide performance guarantees for a greedy algorithm that selects jointly caching retention times and routing decisions.

\subsection{Video Caching}

Due to the popularity of video applications and the large size of the involved files, video content delivery is currently at the focal point of caching research. On the one hand, there are obviously tight delay constraints, especially for streaming services where successive video segments need to be delivered in sync so as to avoid playback stalls. On the other hand, caching decisions are perplexing due to the multiple encoding options. Each video comes in several versions, each of different size, and furthermore the users might have inelastic or elastic needs in terms of video quality. Finally, it is worth mentioning that in live video streaming requests can be predicted with higher precision \cite{valentin}, and this facilitates caching decisions. 

Clearly, video delivery performance is heavily affected by caching. The early work \cite{towsley-video-jsac97} proposed scheduling policies that leverage caching at the user side (buffering) to improve video delivery, and \cite{towsley-video-jsac07} suggested proactive caching of video segments in a peer-assisted CDN. The simplest scenario is that of video on demand (VoD) delivery where one needs to decide which version(s) of each video item to cache \cite{ross-svc}. When the video versions are independent, the caching decisions are only coupled due to the fact that users may have elastic quality needs (hence, the video versions are complementary). When, however, scalable video coding (SVC) is used \cite{schwartz-svc}, additional constraints appear as users can fetch and combine layers of the same video file from different caches. 

The work \cite{poular-infocom14} studies joint video caching and routing in HetNets, aiming to reduce delivery delay and network expenditures, and \cite{poular-infocom16} focuses on delay minimizing policies which are, therefore, also suitable for video streaming. Similarly, \cite{applegate-conext10} formulates a min-cost routing and caching problem for a large VoD network, and \cite{sanchez-svc} analyzes the benefits of SVC for caching networks that deliver video streaming services. More recently, the focus has been shifted to wireless networks with proposals for collaborative video caching \cite{dai-collabor-jsac}, joint routing and caching \cite{chakareski-jsac16}, or network-coding assisted caching of video solutions \cite{ostovari-mass13}.

Reference \cite{video-choi} studies mobile video delivery through D2D links. A base station seeds the devices with videos of possibly different encoding quality, and nearby devices collaborate by exchanging these files. This is formulated as a dynamic problem that maximizes the time-average video quality, through optimized caching and D2D scheduling policies. 

Similarly, \cite{video-ge} studies HTTP-based live streaming services where mobile devices request ultra-high video quality. In these services the end-users often have deteriorated Quality-of-Experience due to the employed congestion control mechanisms in TCP. The authors propose a scheme which performs context-aware transient holding of video segments at the cache-enabled mobile edge which eliminates buffering and substantially reduces initial startup delay and live stream latency.

\subsection{Caching Economics}

The economics of caching is perhaps one of the least explored research areas, rapidly gaining momentum due to the advances in virtualization, that enhance the flexibility in managing  storage resources. Prior work can be broadly categorized to: \emph{(i)} caching cooperation mechanisms, and \emph{(ii)} pricing methods. 

\subsubsection{Cooperation Mechanisms} 
In previous works, e.g., see \cite{korupulu-hierarchical} and \cite{borst-2010}, the term ``cooperative caching'' was used to describe systems where content requests are served by any cache belonging to a set (or, network) of caches. These works, however, take for granted the cooperation among CDNs, mobile operators and users. In practice, these self-interested entities will share their storage resources and coordinate their caching policies, only if they will benefit from cooperation. Prior work shows that if the \emph{incentive alignment} problem is not solved, caching systems experience significant performance loss \cite{papadimitrious-selfish-caching}. Later, \cite{crowcroft-ToN17}  proposed a cooperation mechanism (for a general CN) based on the Nash Bargaining Solution. The latter is attractive as it disperses the cooperation benefits proportionally to the performance each entity would have achieved under non-cooperation. A different suggestion is to use pricing where co-located caches pay for the content they receive from each other, e.g., \cite{dai-collabor-jsac}.

Incentives may be also offered to users in order to assist the network. For example, \cite{biswas-social-caching} discusses the problem of incentivizing users to exchange content by leveraging D2D communications. In a different example, \cite{poular-TNSM-16} proposed a solution where an operator can lease storage and wireless bandwidth of residential access points. Such solutions that involve user equipment are important as their benefits scale with the demand. 

A business model is proposed in this Issue \cite{economics-renzo}, where a Mobile Network Operator (MNO) leases its edge caches to a Content Provider (CP). The latter aims to offload user requests at the edge by maximizing the edge cache hit-ratio with the minimum possible leasing cost. The authors introduce an analytical framework for optimizing CP decisions, which are conditioned on the user association policy of the network. This is an increasingly relevant scenario and follows proposals for deploying edge storage resources at mobile networks, namely at the EPC or base stations. 

\subsubsection{Pricing mechanisms}

The caching economic ecosystem is complex as it includes: payments from Content Providers (CP) to CDNs for content delivery, from CDNs to ISPs for bandwidth and in-network storage, and from users to CPs and ISPs. Pricing employed by the CDN affects how much content the CP places at the edge, which in turn impacts ISP costs and user-perceived performance. The work \cite{tuffin-cdn} studies revenue-maximizing CDN policies, while \cite{towsley-cashing-in} proposes a flexible CDN pricing method. It was shown in \cite{chuang-CDN} that a revenue-seeking cache owner should offer both best effort and guaranteed content delivery services. On the other hand, \cite{economides} has shown that ISPs can increase their marginal profits by imposing data plan caps to users, thus inducing CPs to charge the users with lower prices. These interactions are further perplexed by the new business models such as Telco-CDNs, CP-CDNs, or elastic CDNs \cite{Junho}. 

Finally, it is crucial to make the distinction between \emph{popular} content items (that typical caching algorithms consider), and \emph{important} items that yield higher revenue. For example, \cite{massoulie-utility-cache} models caching as a utility maximization problem instead of cache hit-ratio or delay optimization problem. This allows us to capture the different importance (and hence price) of each content item. Going a further step, \cite{tadrous-Ton-Caching} proposed dynamic pricing and content prefetching techniques for wireless networks, assuming that operators directly charge the end-users for content access. 

%% file: Section-Open-Issues-1Ver12.tex
\section{Open Issues in Caching}\label{section:open-issues}

In this Section we present a set of important open problems in caching. We first discuss representative state-of-the-art caching systems and the challenges they bring. The solution of these problems, clearly, is of high priority and motivates certain research directions that we further analyze.

\subsection{Notable Existing Caching Systems} \label{section:notable-systems}

\subsubsection{Akamai Intelligent Platform} Akamai owns one of the largest CDNs,  delivering today 20$\%$ of the Internet traffic. The 216K caching servers of its \emph{intelligent platform} \cite{sitaraman} are dispersed at network edges (Points-of-Presence, PoPs) offering low-latency (1-10msec) content access around the globe. Several technical challenges arise in such large delivery platforms. First, it is necessary to protect websites from  Distributed-Denial-of-Service attacks \cite{ddos}, and this need motivates the development of caching and filtering techniques that can deal with large volume of requests. Second, the idea of \emph{deep} (or, edge) caching in PoPs improves the CDN performance but reduces user demand per cache, and hence makes the file popularity at the local level highly volatile \cite{deepcaching}. This requirement drives research on the open problem of edge caching, where the goal is to achieve a high hit ratio in caches placed very close to demand (end-user).

Finally, Akamai, among others, uses the idea of cloud or \emph{elastic} CDN where storage resources are dynamically adapted to meet demand \cite{juniper}. This architecture couples storage deployment and caching decisions. Hence, it renders imperative the efficient design of joint storage allocation and content caching policies, and also gives rise to new business models for content caching.

\subsubsection{Google} The Google Global Cache (GCC) system comprises caches installed at ISP premises. The goal of GCC is to serve locally requests for YouTube content, reducing this way off-network bandwidth costs \cite{ggc}. This system grew substantially after YouTube adopted https traffic encryption in 2013. The importance of GCC motivates the study of peering relations between content providers and network operators, and in particular the design of pricing models for leasing in-network caching capacity at operator premises. Another related challenge is security; prior work has proposed schemes for caching with content confidentiality \cite{leguay}, which allows transparent caching for encrypted flows, however the topic of caching encrypted content is still one of the challenging open topics.  

\subsubsection{Netflix Open Connect} Similarly to GCC, the Netflix CDN is partially deployed within ISPs \cite{netflixcom}. However, Netflix video caching faces different challenges from Youtube, mainly because its catalogue is much smaller and the file popularity more predictable. As such, Netflix has been very innovative in studying  spatio-temporal request profiles, popularity prediction mechanisms, and mechanisms to preload the caches overnight and reduce the daylight traffic footprint. An open research challenge in this context is that of early detection of popularity change and online classification of video files as to whether they are cache-worthy or not.

\subsubsection{Facebook Photo CDN} Facebook uses its own hierarchical CDN for delivering pictures to its users. The system leverages web browser caches on user devices, edge regional servers, and the origin caches \cite{huang-facebook}. Notably, browser caches serve almost $60\%$ of traffic requests, due to the fact that users view the same content multiple times. Edge caches serve $20\%$ of the traffic (i.e., approximately $50\%$ of the traffic not served by browser caches), and hence offer important off-network bandwidth savings by locally serving the user sessions. Finally, the remaining $20\%$ of content requests are served at the origin, using a combination of slow back-end storage and a fast origin-cache. The information flow in the Facebook network involves the generation and exchange of content among users, which is the prototypical example of ICN systems. It is therefore of interest to study how ICN caching techniques can improve this architecture.

\subsubsection{Amazon AWS} Part of AWS is the Amazon Cloudfront, a virtual CDN  which utilizes the cloud storage to provide CDN services. Storing 1TB is priced at \$20 \cite{cloudfront_pricing}, and Amazon allows one to dynamically rent caching resources by changing the storage size every one hour. This cloud or elastic CDN architecture, along with similar solutions proposed by Akamai and others, motivate further research on the arising business models, as well as on the dynamic cache placement and dimensioning. 

\subsubsection{Cadami} The Munich-based startup Cadami was the first to implement and evaluate coded caching in a real system \cite{cadami}. The company demonstrated live streaming to $30$ nodes, producing wireless transmission gains of $\times 3$ with realistic wireless channels, file subpacketization, and coding overheads. The most promising applications for such solutions are entertainment systems in long-haul flights, and satellite broadcast systems, and call for further research in coded caching, a topic well-represented in this special Issue.

\subsubsection{3GPP Standards}
Employing caching in wireless networks has been discussed and proposed by many companies in the scope of 3GPP. For example, T-DOC R3-160688 proposes to place an edge cache at an LTE base station either embedded in eNodeB or standalone. T-DOC R3-160828 explores the benefits of local caching. In the scope of 5G-RAN R.14, the report 3GPP TR 36.933 (published in 03-2017) describes the different caching modules that are included in 5G base stations. These standardization efforts pave the road for further research in wireless caching.




\subsection{Caching and Cloud Computing}

\begin{figure}
	\centering
	\includegraphics[width=0.85\linewidth ]{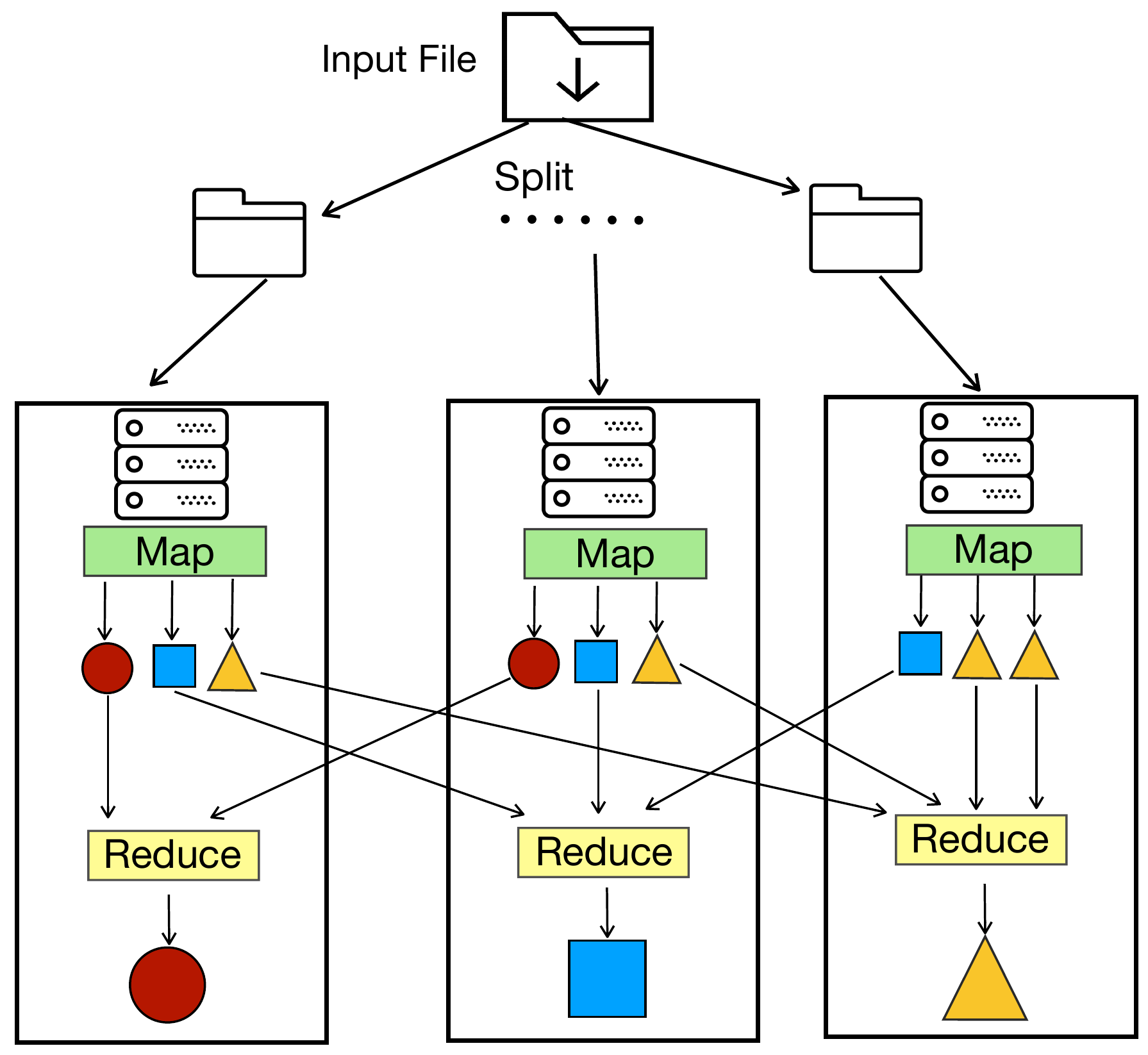}   	
	\caption{\small An example of coded distributed computing. }
	\label{fig:map-reduce}    
\end{figure}

Caching techniques are expected to play an  important role in the upcoming cloud-based network architectures. We describe below two  research directions on the topic of cloud memory.

\subsubsection{Coded Distributed Computing} 
In 2004, Dean and Ghemawat (Google) proposed \emph{map-reduce}, where a large-scale computing task is broken into simple processing of key/value pairs and assigned to parallel processors in an interconnected cluster. The map-reduce paradigm uses the \emph{map} phase to assign processing tasks, and the \emph{reduce} phase to combine answers in order to produce the final result, significantly reducing the total computation time of certain operations (e.g., matrix inversion). The idea of \emph{coded distributed computing} \cite{cdc} is to use coded caching on the reduce step. Combined with careful task assignment, it can dramatically decrease the communication bandwidth consumed during the {reduce} phase. This approach essentially allows us to trade-off node storage with link bandwidth, and can thus accelerate the network-limited map-reduce systems. An example is shown in Fig. \ref{fig:map-reduce}. This recent finding reveals a hitherto hidden connection, or \emph{interoperability}, among bandwidth, storage, and processing; and creates new possibilities for improving network performance through their joint consideration. 

\subsubsection{Caching and Virtualization}
The network virtualization techniques continue to gain momentum. For example, SDN and NFV are considered enablers of cloud-based networks. In this context, CDNs are also expected to migrate to clouds. In particular, caching functionality will be implemented as a Virtual Network Function (VNF) by means of software. Caching VNFs will provide a very flexible environment; they will be instantiated, executed, scaled and destroyed on the fly. Allocating resources for cache VNFs falls into the general framework of network slicing  \cite{algo_slice}, with some special constraints. For example, populating a cache is time-demanding and bandwidth-consuming. More importantly, caches do not satisfy flow conservation; the incoming traffic is partially served by the cache, and only a fraction of traffic continues towards the origin server. Specifically, the larger the caching resource of the VNF, the greater the flow compression. Therefore, VNF embedding for caching must be generalized to include flow compression/decompression constraints \cite{secci}. These new considerations call for a  generalization of the available theory for caching networks.

\subsection{Caching in 5G Wireless Networks and Beyond} 

In the wireless domain, the interoperability of caching, computing and communication techniques opens exciting research directions. Caching trades scarce wireless communication bandwidth and transmission power with (the more cost-effective) memory storage by introducing traffic time-reversal into the system. Caching also enables edge computing capabilities by pre-installing necessary computing software and datasets at the wireless edge nodes. As such, investigating the interplay between these resources is essential for the development of future wireless networks, and several important research questions in this area have been already identified. 

\subsubsection{Performance Characterization of Cache-enabled Wireless Networks}
The first question is information-theory oriented and is related to defining and characterizing the performance limits of cache-enabled wireless networks. In traditional wireless networks, the transmission rate has been a universal performance metric, expressed as a function of signal-to-noise ratio. In the emerging cache-enabled wireless networks, due to the additional memory resource, which varies in size and location, previously adopted performance metrics have become diversified. They include hit probability \cite{eggiovanidis}, \cite{[Bastug-WCN15]}, delivery rate \cite{[Ji-D2D-15]}, \cite{[Bastug-WCN15]}, delivery latency \cite{Dimakis:femtocaching}, \cite{[Sengupta-ciss16]}, \cite{[Xu-TIT17]}, \cite{simeone-d2d-aided-2018}, and traffic load \cite{nielsen}. Whether we need a universal metric that can capture, in a satisfactory fashion, the performance as a function of multi-dimensional resources is a question worth investigating. And if the answer is affirmative, we may need to expand the classic Shannon-type network information theory to study its limiting performance. 

\subsubsection{Tools for Wireless Optimization}
The second research direction concerns the development of efficient and effective algorithms for the optimization of cache-enabled wireless networks. The joint optimization of cache placement and physical layer transmission is often NP-hard \cite{Dimakis:femtocaching} and involves mixed time-scale optimization \cite{[Liu-TSP13]}. This makes these problems particularly challenging to solve, even more since their scale is typically very large and they need to be solved very fast (for enabling dynamic decisions). Furthermore, in wireless networks there are often multiple (collaborating) caches in range of the users, e.g., in multi-tier HetNets, and many possible paths connecting them. Despite the many efforts for designing algorithms that can solve these combinatorial problems (e.g., dual methods, pipage rounding, randomized rounding, etc.), practical challenges prohibit currently the application of these techniques in real systems and further progress needs to be made.

\subsubsection{Support of Emerging Wireless Services}
Finally, another key research thread is to explore the interplay between caching, computing and communications to boost future emerging wireless services, such as mobile AR/VR applications and \emph{vehicle-to-everything} (V2X) communications. These services will often rely on cooperative caching and this raises additional technical questions. Namely, in multi-access caching, finding the route to nearest content replica is a practical challenge, since these services have very limited tolerance in route discovery delay. Therefore, it is important to simplify routing decisions and design them jointly with content discovery. Another interesting aspect is that these services often involve multicast or broadcast transmissions which can greatly benefit from caching. For example, delayed broadcast is currently implemented with parallel unicast sessions, but could be more bandwidth-efficient if caching is employed.

\subsection{Caching with Popularity Dynamics}

\begin{figure*}[t!]
	\centering
	\subfigure[Impact of Time on Popularity]{
		\includegraphics[width=0.24\linewidth ]{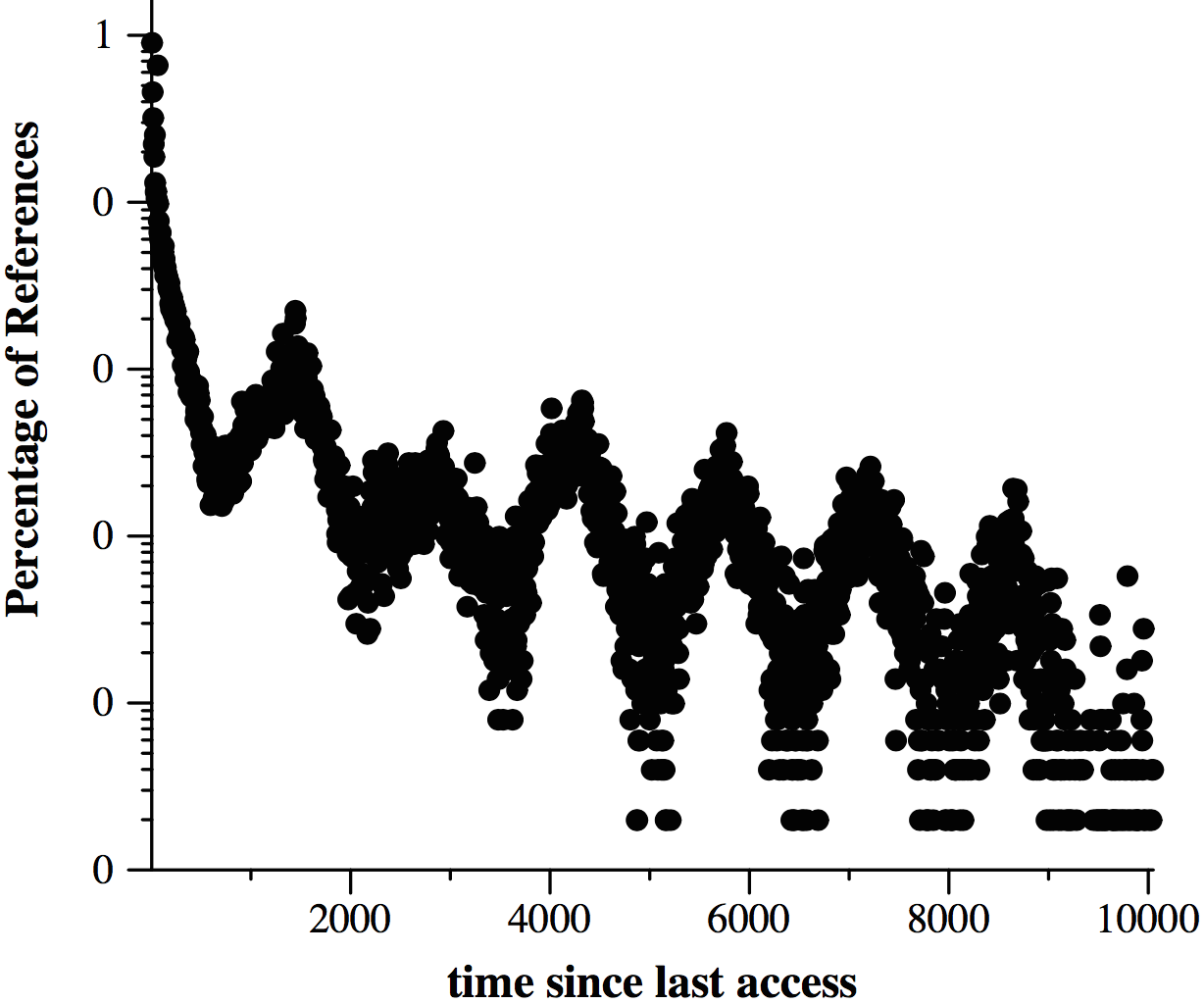}\,\,\,\,\,\,\,\,\,
		\label{fig:popularity1}}
	\subfigure[Popularity Distribution]{
		\centering
		\includegraphics[width=0.22\linewidth]{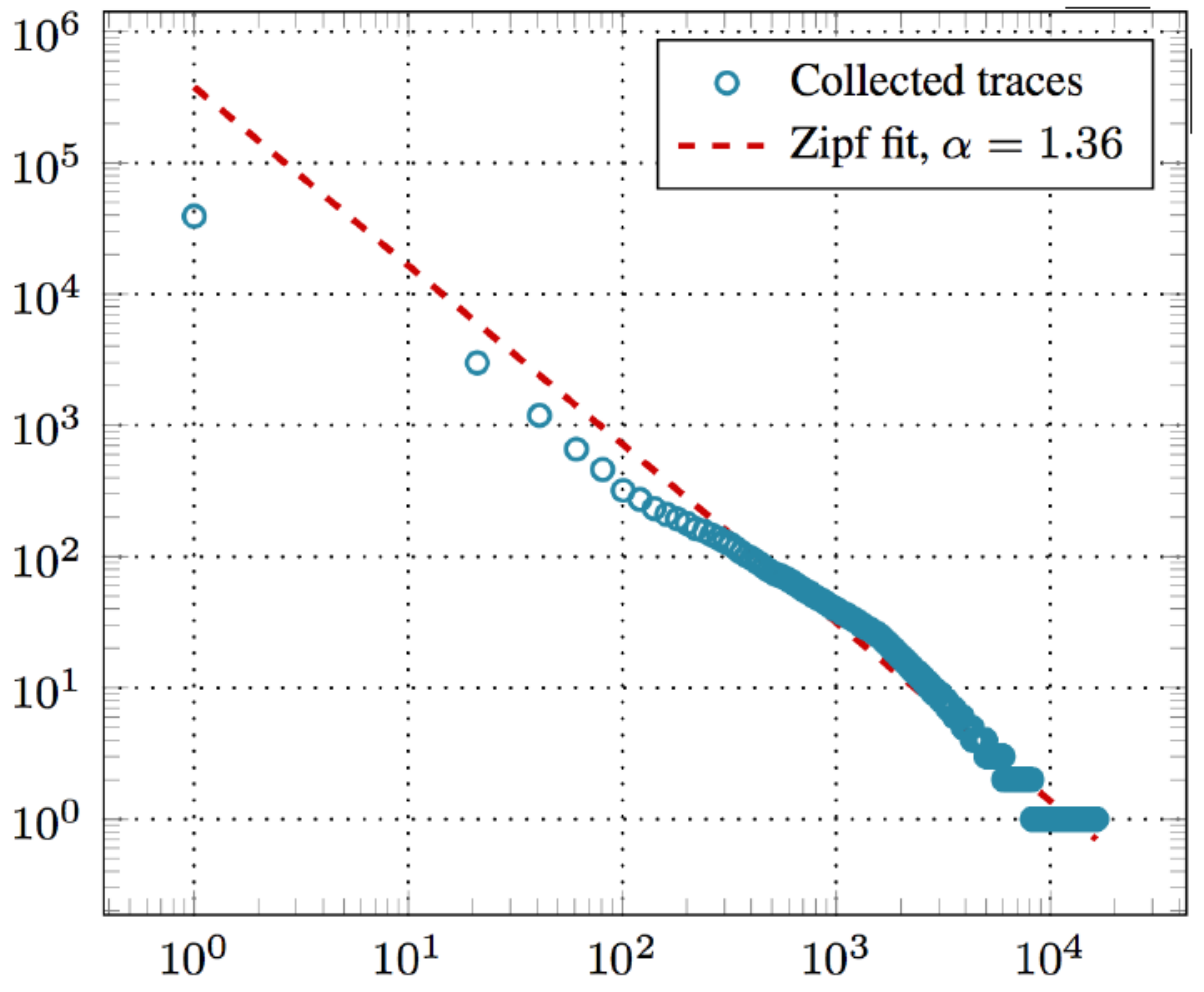}\,\,\,\,\,
		\label{fig:SNM}}
	\subfigure[Poisson Shot Noise Model]{
		\centering
		\includegraphics[width=0.36\linewidth]{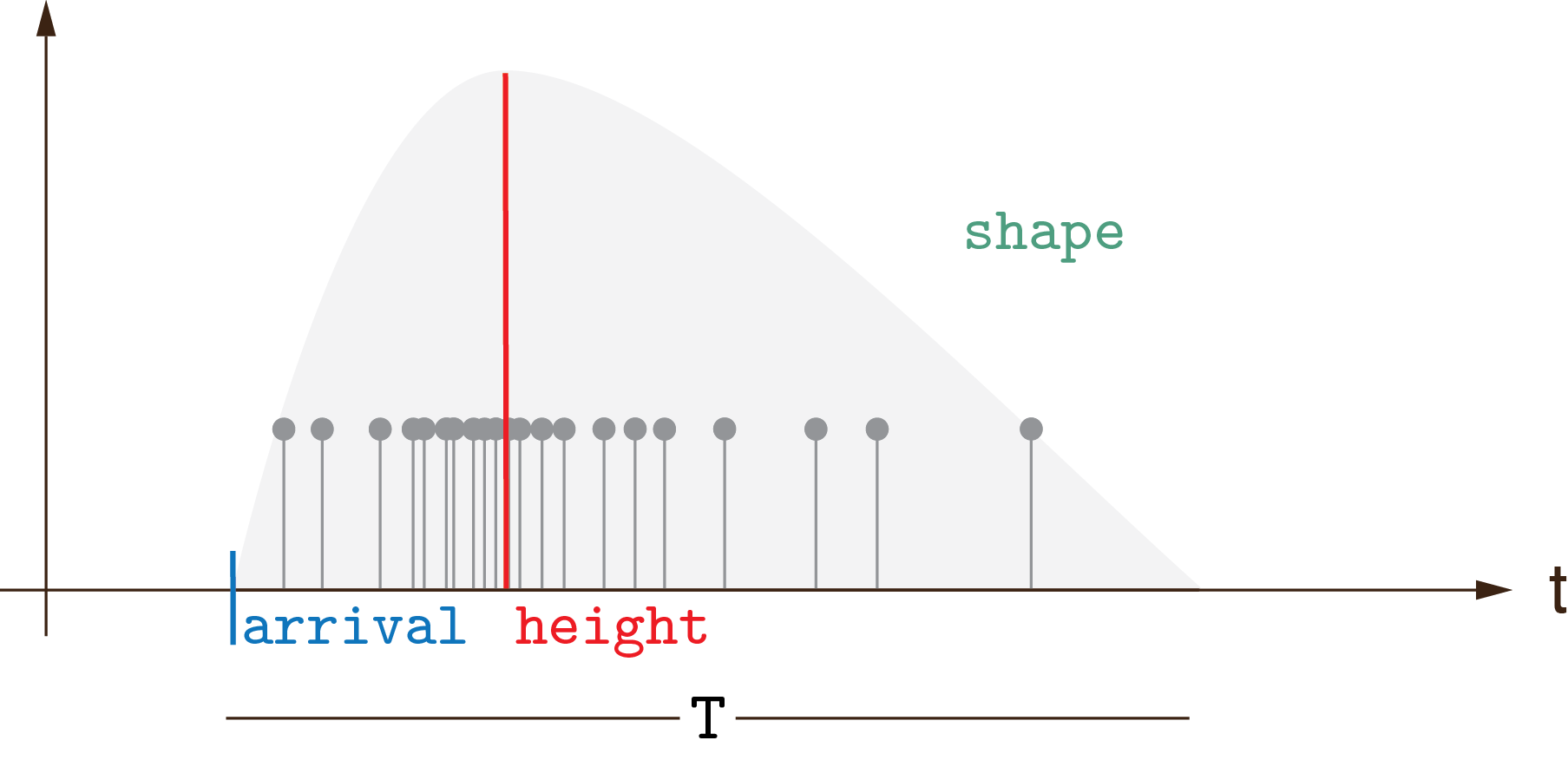}\,\,\,\,\,\,\,
		\label{fig:popularity2}}    	
	\caption{\small  (a):Percentage of references as a function of the time since last access to the same document by the same user (log-log scale), from \cite{cao97}. (b): http requests in mobile operator \cite{Ejder15} (2015); (c): Poisson Shot noise model \cite{traverso}.}
	\label{fig:popularity}    
\end{figure*}

Understanding content popularity is essential to cache optimization; it affects the deployment of caching networks and the design of caching policies, shaping to large extent the overall network performance. In fact, the very notion of diverse popularity of the different content items is what motivated the idea of caching in the first place: ``cache popular items to create a big effect with a small cache''. Yet, understanding, tracking, and predicting the evolution of file popularity in real world is complex and, often, misinterpreted. The community is actively seeking answers to these questions.

\subsubsection{Accurate Popularity Models}
A large body of  work in the literature employs the well-known \emph{Independent Reference Model} (IRM),  which assumes that the content requests are drawn in an i.i.d. fashion from a given distribution. Admittedly, IRM leads to tractable caching models but often at the expense of accuracy. For example, using IRM, we can draw power-law samples in an i.i.d. fashion to  depict quite accurately the request in a real system during a short time interval. Indeed, the power-law  models have been shown to characterize very accurately the popularity within a short time frame \cite{breslau}, i.e., in an interval when popularity can be assumed fixed. However, content popularity is in reality far from stationary, and might change significantly even over few hours. For example, requests of Wikipedia articles have a rapid day-to-day change in popularity rank: half of the top 25 contents change from day to day \cite{hasslinger17}. In Fig.~\ref{fig:popularity1} the phenomenon of ``temporal locality'' is demonstrated, where recently requested contents tend to be more popular \cite{cao97} (note the decrease in the request rate envelope). \emph{In summary, applying IRM to a large time-scale analysis is clearly problematic.}

The importance of content popularity dynamics is reflected in the proliferation of online caching policies such as LRU and LFU, which attempt to adapt caching decisions to temporal locality and popularity fluctuations. These policies do not necessarily provide the best performance, but they are championed in practical engineering systems because they capture some aspects of non-stationarity and they are easy to implement. These policies are often analyzed with stationary popularity or adversarial models. For example, LFU is optimal under IRM (it converges to ``cache the most popular''), LRU over IRM can be analyzed with the characteristic time approximation \cite{che}, and LRU has optimal competitive ratio when the requests are chosen adversarially \cite{sleator85}. Only recently a number of works have studied the performance of dynamic policies with non-stationary popularity. In \cite{traverso}, a inhomogeneous Poisson model was proposed to capture non-stationary popularity, called the Poisson Shot Noise (PSN) model. Under PSN, the LRU performance is provided in \cite{Ahmed}, while \cite{leconte16} gives the optimal policy called ``age-based threshold'' which takes into account the frequency and the age of a content. However, a problem with PSN is that it has too many degrees of freedom, making it quite cumbersome for fitting to real data and optimizing caching systems. The quest for the right non-stationary model is still open.

\subsubsection{Content Popularity Prediction}

Rather than following such reactive techniques, a recent research trend aims to predict content popularity and then optimize accordingly the content placement. For example several past papers look at how a trending file will evolve \cite{Tatar14}, or how social networks can be used to predict the file popularity \cite{Asur10}. More recently, several machine learning techniques have been proposed to assimilate popularity changes in the best manner, namely bandit models \cite{Blasco15}, recommendations \cite{iordanis-caching}, Q-learning \cite{giannakis}, transfer learning \cite{Bastug15}, etc. However, due to its non-stationary nature,  popularity is not easily predicted. In this SI alone, there were 14 submissions on this topic, which reflects how inspiring this challenge is, but also how many different  viewpoints are taken on this subject. We mention here some practical challenges: \emph{(i)} apart from the content popularity, the catalogue of contents is also evolving, \emph{(ii)} the learning rate depends on the volume  of observed samples, and consequently on the aggregation layer  of the studied cache. Learning the popularity at the edge is thus very challenging, \emph{(iii)} the content popularity depends  on the user community characteristics, and geographical clustering of caches has the potentially to improve learning \cite{leconte16}.

%% file: Section-Open-Issues-2Ver12.tex
\subsection{Cooperation, Incentives, and Pricing}

As the caching ecosystem grows more complex, it becomes imperative to align the interests of the key stakeholders so as to alleviate market inefficiencies. Indeed, similarly to other networking areas, is also true that in caching many technical issues can be solved with economic mechanisms. 

\subsubsection{Pricing Cached Content and Elasticity}

User demand often exhibits elasticity that the network can exploit to improve the services and reduce content delivery costs. Users, for example, can often delay their requests and download large content files during off-peak hours, or can use non-congested network paths (e.g., Wi-Fi links). Moreover, the users can submit their content requests in advance so as to assist the network in serving them proactively \cite{tadrous-Ton-Caching}. They can even adapt their requests, e.g., selecting a lower video quality or an already cached video \cite{iordanis-caching}. There are two important open questions here: how to better exploit this elasticity so as to maximize caching performance (or minimize costs) and how to incentivize users to comply accordingly. 

These questions open the discussion about \emph{smart pricing} techniques for cached content that extend beyond managing network congestion \cite{carlee-sdp}. There is clearly an opportunity to couple caching policies with the importance of each content file, measured in terms of revenue (user payments). First steps towards this direction have been made, e.g., see \cite{massoulie-utility-cache}, \cite{spyropoulos-neo} where content popularity is not the sole caching criterion. Charging the cached content delivery in proportion to the induced bandwidth consumption, inversely proportional to its expected cache hit ratio, or based on the service quality improvement it offers to the user, are only some first intuitive suggestions worthwhile investigating.  


\subsubsection{Network and Cache Sharing}

The deployment of infrastructure entails huge capital and operational costs which constitute high market-entry barriers. A solution to this problem is to virtualize and share storage resources, e.g., different CDNs can jointly deploy and manage edge servers. These architectures require mechanisms for deciding: \emph{(i)} how much capital each entity should contribute for the shared storage? \emph{(ii)} how to allocate the virtualized capacity? There are (at least) two levels of cooperation: agree to share the physical resources (\emph{virtualized storage}), and share the cached content (\emph{virtualized content}). The latter option brings higher benefits if the CDNs design jointly their caching policies, and this is one of the most interesting open scenarios in this topic.

Furthermore, cooperation of CDNs with ISPs can bring significant performance and economic benefits. Selecting jointly, for example, the server and route for each content request can reduce both the service delay and network congestion \cite{smaragdakis-CCR}. This coordination is expected to yield significant benefits for wireless edge caching where the network state and demand are highly volatile. Yet, we need to explore how this coordination can be realized, meaning we have to study how to solve these joint optimization problems (caching is already a complex one), and how to disperse the benefits to the collaborating CDNs and ISPs. Finally, elastic CDNs create a new set of problems where cache dimensioning and content caching decisions are devised in the same time scale \cite{Junho}. The flexibility of these architectures enables the very frequent update of these decisions, and therefore it is important to optimize long-term performance criteria for a collection of policies (instead of single-policy metrics).

\subsubsection{Incentive Provision for Hybrid Architectures}

User-owned equipment has increasing capacity and can be considered as an effective caching element of the network. The idea of hybrid CDN-P2P systems is an excellent example in this direction where content delivery (e.g., software patches) is assisted by the end-users \cite{peer5-blog}. In future networks this model can deliver even higher benefits (e.g., asymptotic laws of D2D caching) as it transforms negative externalities (congestion) to positive externalities (through users' collaboration). Yet, this solution requires the design of incentive mechanisms for the users, a problem that has been studied for connectivity services  \cite{ioannidis-sigmetrics}, \cite{iosifidis-upn}. Nevertheless, in case of user-assisted caching new questions arise: How to charge for content that is cached at the user devices? How is time affecting the content price (\emph{freshness})? How to minimize the cost users incur when delivering the content? 


%% file: Section-ConclusionsVer12.tex
\section{Conclusions}\label{section:conclusions}

Caching techniques have a central role in future communication systems and networks. Their beneficial effects are expected to crucially impact core parts of our communication infrastructure, including clouds, 5G wireless systems, and Internet computing at large. At the same time, the ecosystem of caching is ever-changing, constantly requiring ideas for novel architectures and techniques for optimizing their performance. These developments, combined with the recent advances in the domain of resource interactions between storage, bandwidth and processing, create a fascinating research agenda for the years to come.